\documentclass[12pt,preprint]{aastex}

\usepackage{natbib}
\usepackage{epsfig}

\shorttitle{Distribution of DLAs in a $\Lam$CDM Universe}
\shortauthors{Nagamine et al.}

\def\avg#1{\langle#1\rangle}
\newcommand{\Lam}{\Lambda}
\newcommand{\lam}{\lambda}

\newcommand{\mpc}{{\rm Mpc}}
\newcommand{\kpc}{{\rm kpc}}

\newcommand{\kms}{\,\rm km\, s^{-1}}
\newcommand{\Msun}{M_{\odot}}
\newcommand{\Lsun}{L_{\odot}}

\newcommand{\hinv}{h^{-1}}
\newcommand{\himpc}{\hinv{\rm\,Mpc}}
\newcommand{\hikpc}{\hinv{\rm\,kpc}}
\newcommand{\himsun}{\,\hinv{\Msun}}

\newcommand{\ltsim}{\lesssim}
\newcommand{\gtsim}{\gtrsim}
\newcommand{\beq}{\begin{eqnarray}}
\newcommand{\eeq}{\end{eqnarray}}
\newcommand{\dd}{{\rm d}}

\newcommand{\Mdla}{M_{\rm DLA}}
\newcommand{\Mhalo}{M_{\rm halo}}
\newcommand{\Mpeak}{M_{\rm peak}}
\newcommand{\Mmed}{M_{50}}
\newcommand{\Msev}{M_{75}}
\newcommand{\Mmean}{\avg{M_{\rm DLA}}}
\newcommand{\lgMmean}{\avg{\log M_{\rm DLA}}}
\newcommand{\sdla}{\sigma_{\rm DLA}}
\newcommand{\sdlaphys}{\sigma_{\rm DLA}^{\rm phys}}
\newcommand{\sdlaco}{\sigma_{\rm DLA}^{\rm co}}
\newcommand{\Rab}{{\cal R}_{AB}}
\newcommand{\bphys}{b_{\rm phys}}
\newcommand{\hsev}{h_{70}}
\newcommand{\hsevkpc}{h_{70}^{-1}\,{\rm kpc}}
\newcommand{\rdla}{r_{\rm DLA}}
\newcommand{\Adla}{A_{\rm DLA}}

\newcommand{\Om}{\Omega_{\rm m}}
\newcommand{\Ol}{\Omega_{\Lam}}
\newcommand{\Ob}{\Omega_{\rm b}}

\newcommand{\HI}{H{\sc i}\,\,}
\newcommand{\NHI}{{N_{\rm HI}}}

\newcommand{\esn}{\epsilon_{\rm SN}}
\newcommand{\Mwdot}{\dot{M}_W}
\newcommand{\Msdot}{\dot{M}_{\star}}
\newcommand{\rhothw}{\rho_{\rm decoup}}
\newcommand{\rhoth}{\rho_{\rm th}}
\newcommand{\tdecoup}{t^{\rm max}_{\rm decoup}}
\newcommand{\zetadecoup}{\zeta_{\rm decoup}}
\newcommand{\Lmaxtravel}{\ell_{\rm maxtravel}}

\begin{document}

\title{Distribution of Damped Lyman-$\alpha$ Absorbers in a $\Lambda$ Cold Dark Matter Universe}

\author{Kentaro Nagamine\altaffilmark{1,2}, Arthur M. Wolfe\altaffilmark{1}, Lars Hernquist\altaffilmark{3}, Volker Springel\altaffilmark{4}}

\altaffiltext{1}{Center for Astrophysics and Space Sciences, University of California, San Diego, 9500 Gilman Dr., La Jolla, CA 92093-0424}
  
\altaffiltext{2}{Present address: Department of Physics \& Astronomy, University of Nevada, Las Vegas, 4505 Maryland Parkway, Las Vegas, NV 89154-4002} 

\altaffiltext{3}{Department of Astronomy, Harvard University, 60 Garden Street, Cambridge, MA 02138, U.S.A.}

\altaffiltext{4}{Max-Planck-Institut f\"{u}r Astrophysik,
         Karl-Schwarzschild-Stra\ss{}e 1, 85740 Garching bei
         M\"{u}nchen, Germany}


\begin{abstract}

We present the results of a numerical study of a galactic wind model
and its implications on the properties of damped Lyman-$\alpha$
absorbers (DLAs) using cosmological hydrodynamic simulations.  We vary
both the wind strength and the internal parameters of the the wind
model in a series of cosmological smoothed particle hydrodynamics
(SPH) simulations that include radiative cooling and heating by a UV
background, star formation, and feedback from supernovae and galactic
winds.  To test our simulations, we examine the DLA
`rate-of-incidence' as a function of halo mass, galaxy apparent
magnitude, and impact parameter.  We find that the {\it statistical}
distribution of DLAs does not depend on the exact values of internal
numerical parameters that control the decoupling of hydrodynamic
forces when the gas is ejected from starforming regions, although the
exact spatial distribution of neutral gas may vary for individual
halos.  The DLA rate-of-incidence in our simulations at $z=3$ is
dominated ($80-90$\%) by the faint galaxies with apparent magnitude
$\Rab<25.5$.  However, interestingly in a `strong wind' run, the
differential distribution of DLA sight-lines is peaked at $\Mhalo =
10^{12} \himsun$ ($\Rab \simeq 27$), and the mean DLA halo mass is
$\Mmean=10^{12.4}\himsun$ ($\Rab \simeq 26$).  These mass-scales are
much larger than those if we ignore winds, because galactic wind
feedback suppresses the DLA cross section in low-mass halos and
increases the relative contribution to the DLA incidence from more
massive halos.  The DLAs in our simulations are more compact than the
present-day disk galaxies, and the impact parameter distribution is
very narrow unless we limit the search for the host galaxy to only
bright Lyman-break galaxies (LBGs).  The comoving number density of
DLAs is higher than that of LBGs down to $\Rab=30$ mag if the physical
radius of each DLA is smaller than $5\,\hsevkpc$.  We discuss
conflicts between current simulations and observations, and potential
problems with hydrodynamic simulations based on the cold dark matter
model.
\end{abstract}



\section{Introduction}
\label{sec:intro}

It is widely believed that the galactic wind phenomenon has very
strong effects on the process of galaxy formation.  We observe winds
of hot gas emanating from starburst galaxies such as M82
\citep{McCarthy87}, and also find evidence for large-scale outflows in
the spectrum of high-redshift starforming galaxies such as LBGs
\citep[e.g.][]{Pettini01, Ade03}.  Such galactic winds heat up the
intergalactic medium (IGM) and enrich it with heavy elements.
Therefore in recent years theorists have incorporated phenomenological
models of galactic wind into numerical simulations and studied its
influence on galaxy formation and chemical enrichment of the IGM
\citep[][]{Theuns02, SH03b, Cen05}.  From those studies, it became
clear that the observed metallicity of the Ly-$\alpha$ forest can be
only understood if the effect of galactic winds is considered.  In
this paper, we test the model of galactic wind presently included in
our cosmological simulations against the observations of damped
Lyman-$\alpha$ absorbers (DLAs).\footnote{DLAs are historically
defined to be absorption systems with neutral hydrogen column
densities $\NHI > 2\times 10^{20}$ cm$^{-2}$ \citep{Wolfe86}.}

Within the currently favored hierarchical $\Lam$ cold dark matter
(CDM) model, DLAs observed in high redshift quasar absorption lines
are considered to arise from radiatively cooled neutral gas in dark
matter halos.  They are known to dominate the neutral hydrogen mass
density at high redshift \citep[e.g.,][]{Lanzetta95, Storr00}, and
hence provide a significant reservoir of cold neutral gas for star
formation.  If this picture is correct, then DLAs are closely linked
to star formation and must be located inside or in the vicinity of
galaxies within dark matter halos (hereafter often just `halos').  For
these reasons, DLAs provide an excellent probe of high redshift galaxy
formation that complements the study of high redshift galaxies by the
direct observation of stellar light.  They also provide useful
opportunities to test hydrodynamic simulations of galaxy formation.

Despite significant observational effort over the years
\citep[e.g.,][]{Wolfe86, Lanzetta95, Wolfe95, Storr00, Rao00, Pro01, Pro02,
Peroux03, Chen03b}, the true nature of DLA galaxies (i.e., galaxies
that host DLAs) is still uncertain, and it is unclear how DLAs are
distributed among dark matter halos. The observed large velocity
widths of low-ionization lines support a large, thick disk hypothesis
\citep[e.g.,][]{Wolfe86, Turnshek89, Pro97, Pro98}, while at the same
time there is evidence that a wider range of galaxies could be DLA
galaxies \citep[e.g.,][]{LeBrun97, Kulkarni00, Kulkarni01, Rao00,
Chen03b, Wea05}.  \citet*{Hae98} showed that their SPH
simulations were able to reproduce the observed asymmetric profiles of
low-ionization absorption lines, and argued that DLAs could
be protogalactic gas clumps rather than well-developed massive disks
that have settled down.  However their simulations did not include the
effects of energy and momentum feedback owing to star formation, and
they only analyzed a few systems.

To interpret the observations of DLAs, it is useful to define the
`rate-of-incidence' distribution as a function of halo mass or
associated galaxy magnitude.  The `rate-of-incidence' of DLAs (often
just `DLA incidence'), ${\rm d}N/{\rm d}z$, i.e., the probability of
finding a DLA along a line-of-sight per unit redshift, could either be
dominated by low-mass halos, or by very massive halos.  In the former
case, the DLA cross section in low-mass halos is significant, and
because the halo mass function in a CDM universe is a steeply
increasing function with decreasing halo mass, the net contribution to
the DLA rate-of-incidence from low-mass halos dominates over that from
massive halos. One of the critical elements that needs to be
determined is the DLA cross section as a function of halo mass.

Many authors have used semi-analytic models of galaxy formation based
on the hierarchical CDM model to study the distribution of DLAs.
\citet{Mo98} improved the model of \citet{Kau96} to study the
formation of disk galaxies by following the angular momentum
distribution, and examined the distribution of DLA rate-of-incidence
as a function of halo circular velocity and impact parameter.  
While these calculations provide insights on what is expected
in hierarchical CDM models, they rely 
on assumptions about the mass fraction of disks
and the geometry of the gas distribution relative to those of dark
matter halos.  Furthermore they are not able to directly account for
dynamical effects from violent merging of halos/galaxies and
associated heating/cooling of gas.  \citet{Hae00} studied the
luminosity and impact parameter distribution of DLAs using simple
scaling relationships from both observations and simulations and an
analytic halo mass function.  
There are some recent works that discuss the distribution and 
the physical properties of DLAs based on semi-analytic models of 
galaxy formation \citep{Maller01, Okoshi04, Okoshi05}. 
While these comprehensive semi-analytic models did not require 
assumptions about the cold gas mass fraction in halos, 
they still had to make some choices for the geometry 
of the gas distribution, such as an exponential radial profile 
of H\,{\sc i} column density in the case of \citet{Okoshi04}, 
focusing on the virialized systems.  
We will compare our results in this paper to those of the above 
authors in what follows.

On the other hand, numerical simulations are able to describe
dynamical effects in a more realistic manner than semi-analytic
models, but instead have resolution and box size limitations owing to
finite computational resources.  State-of-the-art cosmological
hydrodynamic simulations that evolve comoving volumes larger than
$\sim (10\,\himpc)^3$ can now achieve a spatial resolution of $\sim$ 1
kpc in comoving coordinates, but have the fundamental problem of being
unable to produce a large population of realistic disk galaxies with
the observed number density at $z=0$ \citep[e.g.,][]{Robertson04}.
This is the so-called `angular momentum problem' where the transfer
of angular momentum from baryons to dark matter is excessive 
and the disks in the simulations become too small relative to the
real ones. Even so, implementations of the physics of star formation 
and feedback have improved over the past several years
\citep[e.g.,][]{SH02, SH03a, SH03b, Cen05}, therefore it is of
interest to compare results with those obtained from simulations that
did not include star formation and feedback \citep{Hae98}. 

In an earlier paper, \citet*{NSH04a} studied the neutral hydrogen mass
density, column density distribution, DLA cross section, and the
rate-of-incidence using a series of SPH simulations with varying box
size and feedback strength.  One of their results was that the DLA
cross section in low-mass halos depends on both resolution and the
feedback strength.  This results in a strong variation 
in the distribution of DLA incidence which was often
neglected in other numerical studies \citep{Katz96b, Gardner97a,
Gardner97b, Gardner01}.  
In this paper we first check that our results are not strongly 
affected by the internal parameters of the wind model that  
control the decoupling of the hydrodynamic force when the 
gas is ejected from a star-forming region. 
Then we examine the DLA rate-of-incidence as functions of 
dark matter halo mass, apparent magnitude of DLA galaxies, 
and impact parameter. 
We compare our results with the observational results 
by \citet{Pro05} that are derived from the 
Sloan Digital Sky Survey (SDSS) Data Release 3.
We have also improved the accuracy of the dark matter halo mass 
function by performing more accurate integral of the power spectrum
when calculating the mass variance $\sigma(M)$ of the density field.


\section{Simulations}
\label{sec:simulation}

We use the {\small GADGET-2} code \citep{Springel05} which
employs the Smoothed Particle Hydrodynamics (SPH) technique.  It
adopts the entropy-conservative formulation of \citet{SH02} which
largely alleviates the overcooling problem that the previous
generation of SPH codes suffered. Our simulations include radiative
cooling and heating with a uniform UV background of a modified
\citet{Haardt96} spectrum \citep{KWH96, Dave99}, star formation,
supernova feedback, a phenomenological model for galactic winds
\citep{SH03b}, and a sub-resolution multiphase model of the ISM
\citep{SH03a}.

We use a series of simulations of varying box size and particle 
number (see Table~\ref{table:sim}) in order to assess 
the impact of numerical resolution on our results. 
Also, the strength of galactic wind feedback is varied 
among the O3 (no wind), P3 (weak wind), and Q3 (strong wind) runs, 
allowing us to study the consequences of feedback on our results.   
The adopted cosmological parameters of all runs are 
$(\Om,\Ol,\Ob,\sigma_8, h)= (0.3, 0.7, 0.04, 0.9, 0.7)$. 
We also use the notation $h_{70} = h / 0.7$, where $h$ is the 
Hubble parameter in units of 100\,km\,s$^{-1}$\, Mpc$^{-1}$. 
In this paper, we only use simulations with a box size of $10\himpc$ 
in order to achieve high spatial resolution,
and focus on $z=3$ since this is one of the epochs 
where the largest observational data are available and hence 
more accurate comparisons to observations are possible. 

\subsection{Dependence on the wind parameters}

The details of the star formation model is given elsewhere
\citep{SH03a, SH03b, NSH04b}, so we will not repeat them here. 
In the galactic wind model adopted in our simulations,  
gas particles are stochastically driven out of the 
dense starforming regions with extra momentum 
in random directions. 
The rate and amount of extra kinetic energy is chosen 
to reproduce observed mass-loads and wind-speeds. 
The wind mass-loss rate $\Mwdot$ is assumed to be proportional to 
the star formation rate, and the wind carries a fixed fraction 
$\chi$ of SN energy:
\beq
\Mwdot = \eta \Msdot, 
\eeq
and
\beq
\frac{1}{2} \Mwdot v_w^2 = \chi \esn \Msdot.
\eeq
A fixed value of $\eta = 2$ is adopted for the wind mass-loss rate, 
and $\chi = 0.25$ (weak wind) \& 1.0 (strong wind) for
the wind energy fraction. 
Solving for the wind velocity from the above two equations, 
the two wind models correspond to speeds of 
$v_w = 242~\kms$ and $484~\kms$, respectively. 
This wind energy is not included in $\esn$ discussed above, 
so in our simulations $(1+\chi)\,\esn$ is the total SN energy 
returned into the gas.

\begin{figure}
\epsscale{1.0}
\plotone{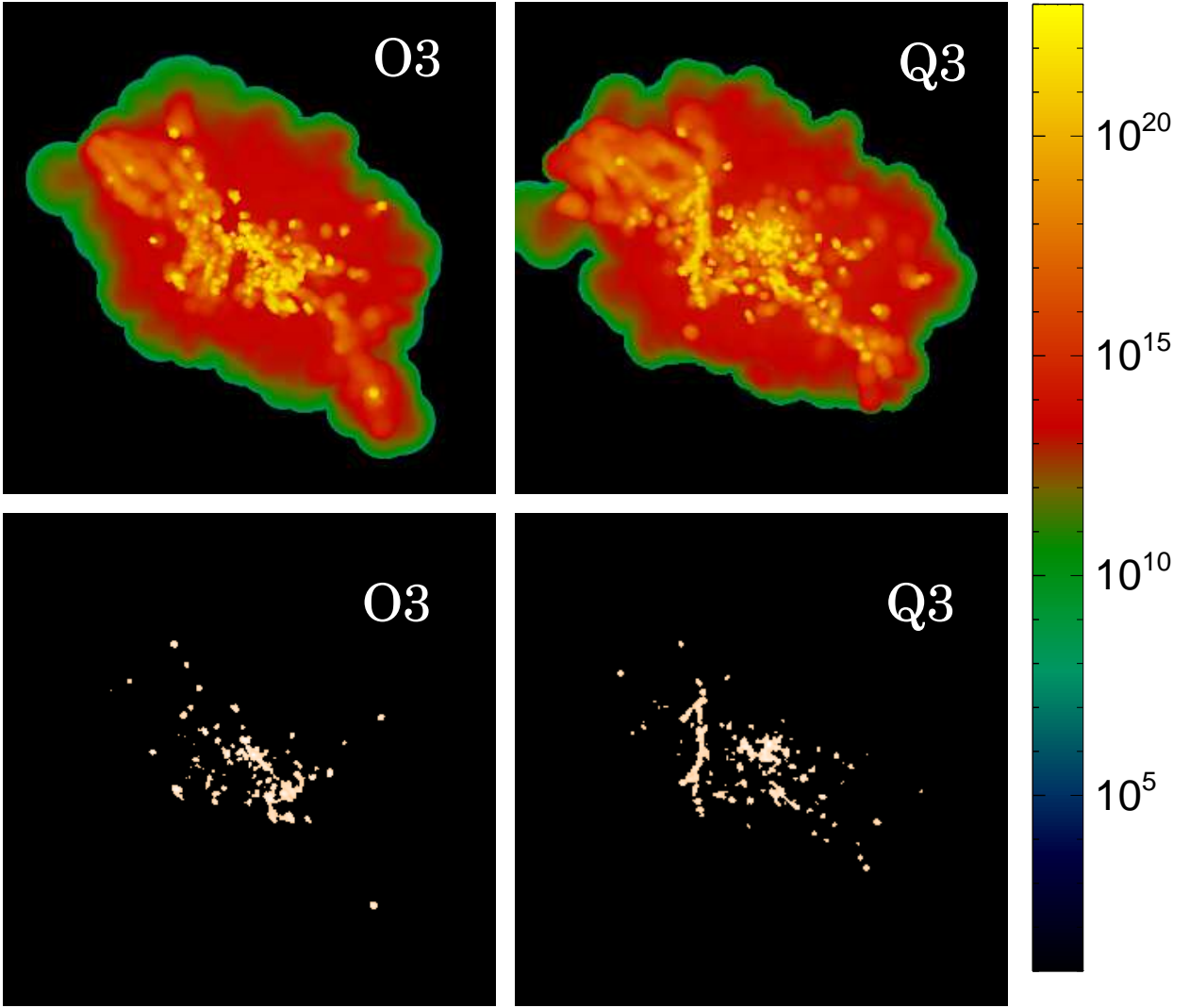}
\caption{Comparison of $\NHI$~[cm$^{-2}$] (top two panels and the range 
of values shown on the side with the color gradient) and 
DLA (bottom two panels) distribution in a dark matter halo of 
a mass $M_{h} \sim 1.7\times 10^{12} \himsun$ at $z=3$ 
for the O3 (no wind) and Q3 (strong wind) run. 
Each postage stamp is about comoving $400 \hikpc$ 
(physical $\sim 100 \hikpc$) on a side.
The DLA cross section in the O3 run is slightly larger than 
that of the Q3 run.
}
\label{fig:compare_OQ}
\end{figure}

Figure~\ref{fig:compare_OQ} compares the distribution of $\NHI$ and
DLAs in a dark matter halo with a mass $M_{h} \sim 1.7\times 10^{12}
\himsun$ at $z=3$ for the O3 (no wind, $\chi = 0.0$) run and Q3
(strong wind, $\chi = 1.0$) runs.  The DLA cross section of this halo
is slightly higher in the O3 run than the Q3 run.  The DLA columns are
more concentrated near the center for the O3 run than the Q3 run,
because without the wind the neutral gas is able to cool and sink
deeper into the potential well.  In the Q3 run, the DLA columns are
distributed more broadly than the O3 run.  In particular in the
western side of the halo, there is a filamentary structure of DLA
columns arising from overlapping DLA clouds.  These DLA clouds may
originate from the gas ejected by the wind but later cooled due to
high gas density within the halo.

When a gas particle goes into the wind mode, our numerical wind scheme
turns off its hydrodynamic interactions for a brief period of time to
allow the particle to escape from the dense star-forming region. This
is done in order to obtain a well-controlled mass-flux in the wind,
which we picture to emanate from the surface of the star-forming
region. Without the decoupling, the mass-loading would be boosted when
the accelerated wind particle kicks and entrains other particles in
the star-forming gas.  The brief decoupling is controlled by two
internal numerical parameters that limit how long the hydrodynamic
forces are ignored.  The primary parameter $\zetadecoup =
\rho / \rhoth $ determines the density $\rho$ in units of the
threshold density $\rhoth$ for star formation that the escaping
wind particle needs to reach before it is allowed to interact normally
again. The idea of letting the wind blow from the surface of the ISM
is therefore described by the condition $\zetadecoup \le 1$. A
secondary numerical parameter $\Lmaxtravel$ is introduced in
order to limit the maximum time $\tdecoup = \Lmaxtravel / v_w$ 
a wind particle may stay decoupled. This length
scale was just added as a precaution against the unlikely -- but in
case of weak winds perhaps possible -- event that a wind particle
`gets stuck' in the ISM, i.e. cannot climb out of the gravitational
potential well far enough to reach the density $\zetadecoup\, \rhoth$. 
We stress that our expectation is that the
outcome of our simulations is rather insensitive to the detailed
choices for these two parameters, something that we have also
confirmed in simulations of isolated galaxies when developing the
model. However, in order to test whether this is also the case in our
much less well resolved cosmological simulations, we have also ran
four test simulations with identical initial conditions as the
original Q3 run but with different values of these two internal wind
parameters, as summarized in Table~\ref{table:windparams}. In
particular, we have varied $\zetadecoup$ by two orders of
magnitude around our default choice of $0.1$, and we changed
$\Lmaxtravel$ between 4 and 100$\,h^{-1}{\rm kpc}$ around
our default choice of $20\,h^{-1}{\rm kpc}$.

\begin{figure*}
\begin{center}
\epsscale{1.0}
\plotone{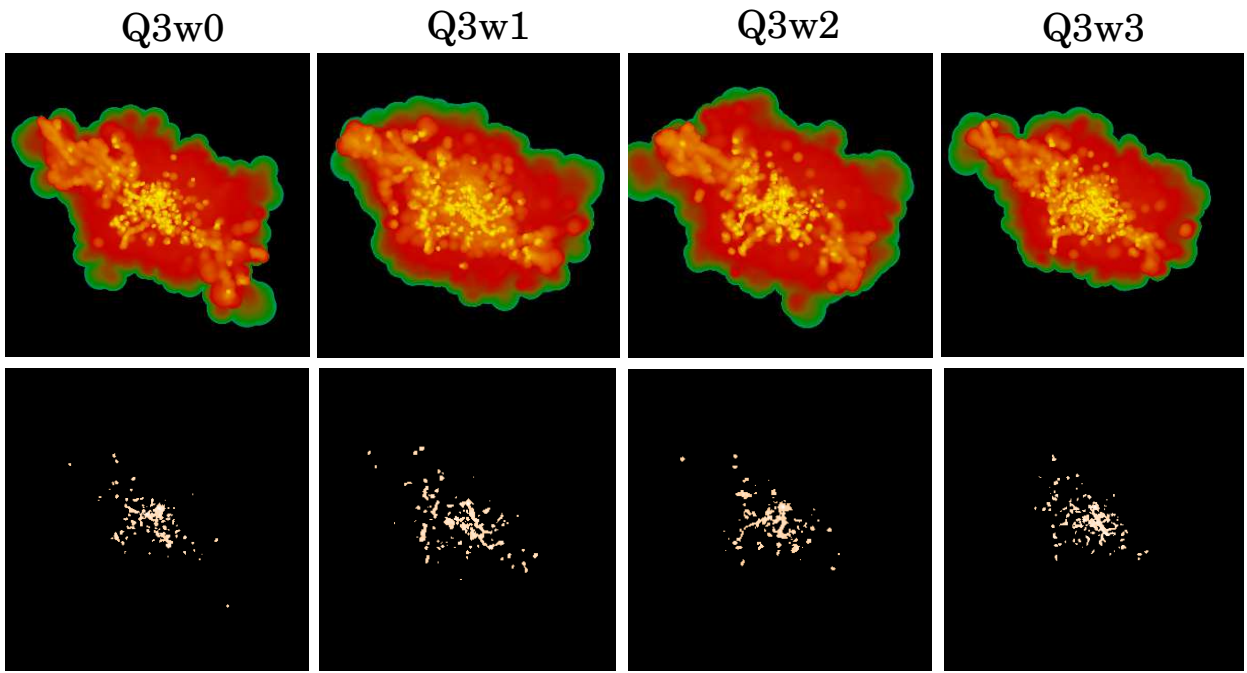}
\caption{Distributions of $\NHI$ (top row)
and DLAs (bottom row) for the same dark matter halo 
as in Fig.~\ref{fig:compare_OQ} in the four test runs 
with the same color scale. 
Each postage stamp is about comoving $400 \hikpc$ 
(physical $\sim 100 \hikpc$) on a side.
}
\label{fig:pics}
\end{center}
\end{figure*}

The panels in Figure~\ref{fig:pics} show the \HI column density
distribution and the DLA distribution for the same halo as in
Fig.~\ref{fig:compare_OQ} in the four test runs.  The overall
distribution of gas is similar between the different runs, but the
exact locations of DLAs in each halo are not identical as a result of
the changes in the decoupling prescription. Some of the difference is
presumably introduced by the randomness involved in the ejection of
the gas particles themselves, so that statistical comparisons between
the results of the different simulations are warranted.

We hence examine the DLA cross section as a function of halo mass to
see whether the distribution of DLAs is {\em statistically} different
or not.  \citet{NSH04a} quantified the relationship between the {\it
total} DLA cross section $\sdlaco$ (in units of comoving
$h^{-2}$\,kpc$^2$; note that Fig.~$2-4$ of \citet{NSH04a} plotted
comoving $\hsev^{-2}\,\kpc^2$) and the dark matter halo mass (in units
of $\himsun$) at $z=3$ as \beq \log\,\sdlaco = \alpha\, (\log\,\Mhalo
- 12) + \beta ,
\label{eq:sigma}
\eeq
with slopes $\alpha = 0.72, 0.79, 0.84, 0.93, 1.02$ and the 
normalization $\beta = 3.94, 3.99, 3.98, 4.03, 4.18$ for the O3, 
P3, Q3, Q4, and Q5 runs. 
The slope $\alpha$ is always positive, and the massive halos
have larger DLA cross section, but they are more scarce compared
to less massive halos. 
The quantity $\beta$ gives the value of $\log\sdlaco$ at 
$\Mhalo = 10^{12}\himsun$.  
Two qualitative trends were noted:
(1) As the strength of galactic wind feedback increases 
(from O3 to Q3 run), the slope $\alpha$ becomes steeper 
while the normalization $\beta$ remains roughly constant. 
This is because a stronger wind reduces the gas in low-mass halos
at a higher rate by ejecting the gas out of the potential well of 
the halo.  
(2) As the numerical resolution is improved (from Q3 to Q5 
run), both the slope and the normalization increase. 
This is because with higher resolution, star formation 
in low-mass halos can be described better and 
as a result the neutral gas content is decreased due to winds. 
On the other hand, a lower resolution run misses the 
early generation of halos and the neutral gas in them. 

\begin{figure}
\begin{center}
\epsscale{1.1}
\plotone{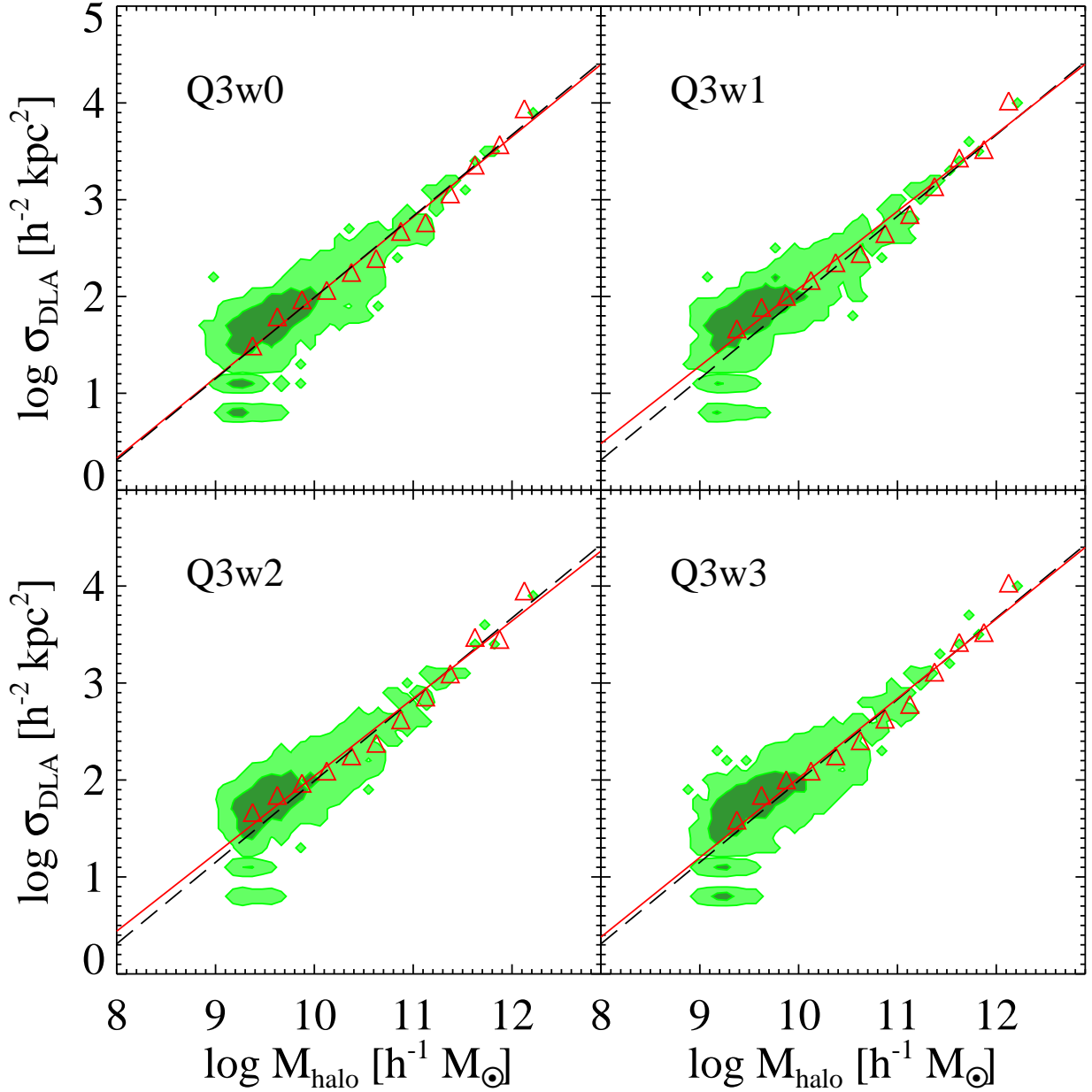}
\caption{DLA cross section vs. halo mass at $z=3$ for the four test runs with different wind parameters. Open triangles are the median in each halo mass bin. The red solid line is the power-law fit to these median points (Table~\ref{table:fitparams}), and the black dashed line is the fit to the original Q3 run. The shaded contour in the background is the actual distribution of halos equally spaced in logarithmic scale. 
}
\label{fig:area}
\end{center}
\end{figure}

Figure~\ref{fig:area} shows the distribution of DLA cross section as a
function of dark matter halo mass for the four test runs with
different decoupling parameters.  Open triangles are the median in
each halo mass bin.  The solid line is the power-law fit to these
median points (see Table~\ref{table:fitparams}), and the dashed line
is the fit to the original Q3 run.  The shaded contours in the
background give the actual distribution of halos equally spaced in
logarithmic scale.  Table~\ref{table:fitparams} shows that the
distributions of DLA cross sections in the four test runs are rather
similar to that of the original Q3 run.  There is a slight tendency
that the slope is shallower in the four test runs than in the original
Q3 run, something that could be related to extensive changes in the
time integration and force calculation scheme of the simulation code
relative to the older code version used in the original Q3 run, rather
than to systematics introduced by the decoupling scheme. In any case,
the fluctuations in our results introduced by drastic variations of
the internal numerical parameters used in the decoupling scheme are
much less than the large differences between the Q3 (strong wind) and
O3 runs (no wind), meaning that the decoupling parameters are not a
significant source of uncertainty in our results.

\begin{figure}
\begin{center}
\resizebox{9.5cm}{!}{\includegraphics{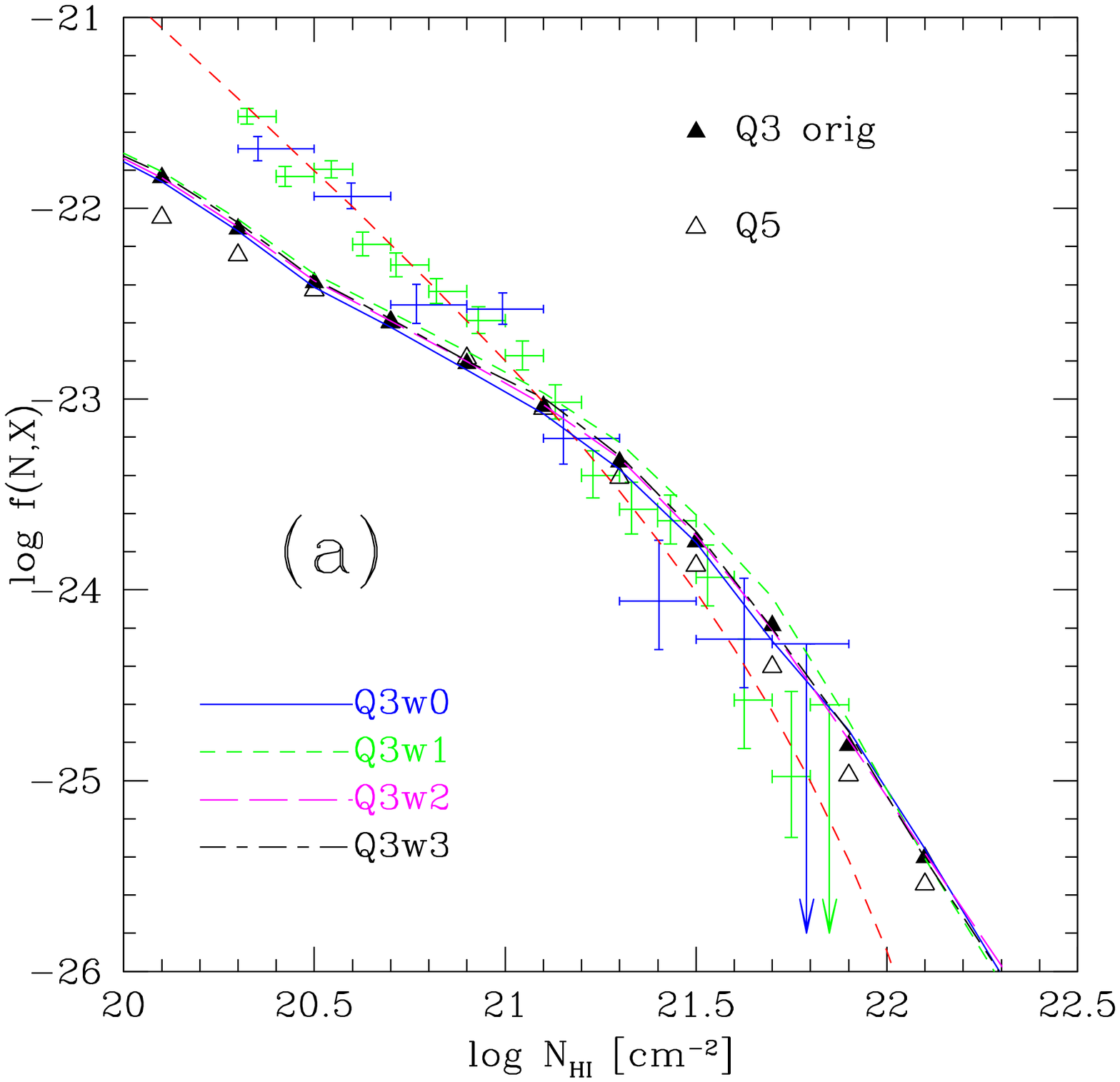}}\\
\vspace{0.3cm}
\resizebox{9.5cm}{!}{\includegraphics{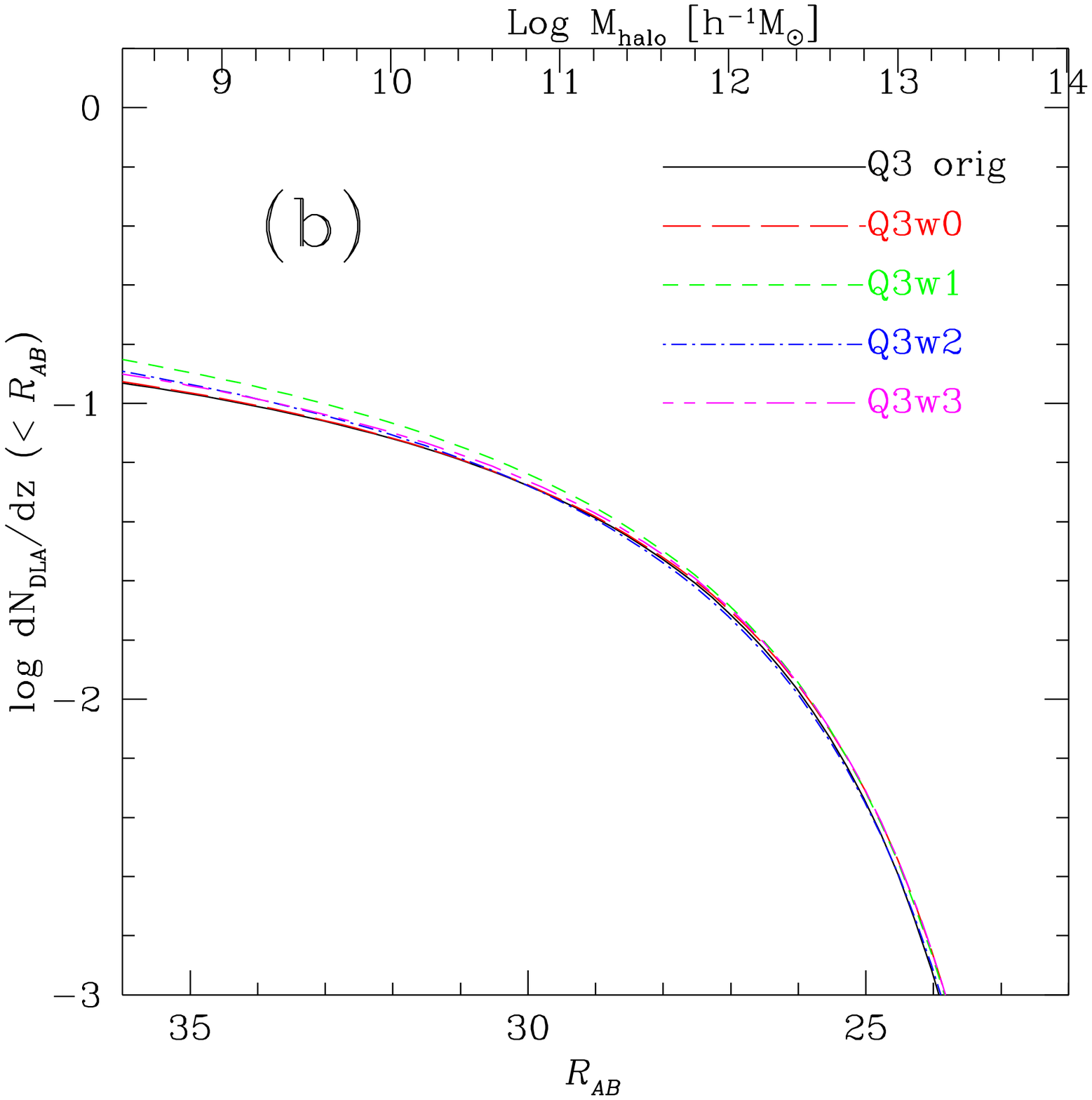}}\\
\caption{{\it Panel (a):} \HI column density distribution function at $z=3$ for the four test runs with different wind parameters. The data points and the red short-dashed line are from \citet{Pro05}. The solid and open triangle is the results from the original Q3 and Q5 runs. 
{\it Panel (b):} Cumulative rate-of-incidence as a function of $\Rab$ magnitude for the test runs and the original Q3 run.  The results of the test runs are not so different from the original Q3 run.  
}
\label{fig:coldist}
\end{center}
\end{figure}

Figure~\ref{fig:coldist} shows the \HI column density distribution
function and the cumulative rate-of-incidence for the four test runs
\citep[see][for the definition of these two quantities]{NSH04a}.
These figures also show that the statistical distribution of DLAs in
the four test runs is similar to the original Q3 run.  Likewise, the
impact parameter distribution which we will discuss in
Section~\ref{sec:impact} is also very similar to the original Q3 run
for the four test runs.  Therefore we focus on the original O3, P3,
Q3, and Q5 runs in the following sections.


\section{Differential distribution of rate-of-incidence as a function of halo mass}
\label{sec:rate}

\begin{figure}
\begin{center}
\resizebox{14cm}{!}{\includegraphics{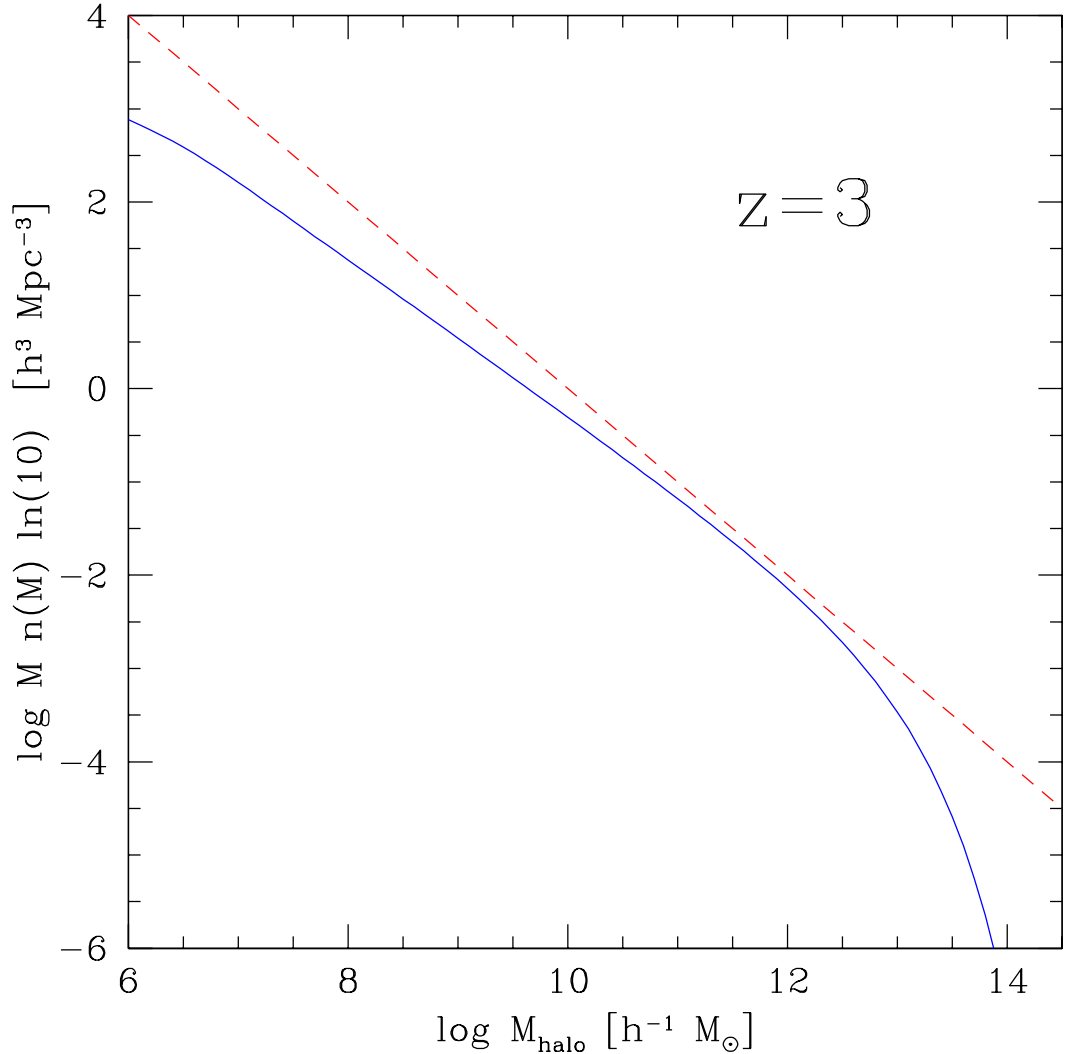}}
\caption{Dark matter halo mass function of \citet{Sheth99} at $z=3$,  
shown in the form of $M\cdot n(M)$ to ease the interpretation of 
Equation~(\ref{eq:rate}). The red dashed line shows a power-law 
$n(M)\propto M^{-2}$. We use the transfer function of \citet{E-Hu} 
and the power spectrum is normalized to $\sigma_8 = 0.9$. 
}
\label{fig:st}
\end{center}
\end{figure}

The differential distribution function of DLA incidence 
can be computed as
\beq
\frac{\dd N_{\rm DLA}}{\dd z\, \dd\log M} = 
\frac{\dd r}{\dd z} [\, M\, n(M,z)\, \ln(10)\,]\, \sdlaco(M,z),
\label{eq:rate}
\eeq where $n(M,z)$ is the dark matter halo mass function, and $\dd
r/\dd z = c/H(z)$ with $H(z)=H_0 E(z) = H_0\sqrt{\Om(1+z)^3+\Ol}$ for
a flat universe.  We use the \citet{Sheth99} parameterization for
$n(M)$ as shown in Figure~\ref{fig:st}.  Note that the dependence on
the Hubble constant disappears on the right-hand-side of
Equation~(\ref{eq:rate}) because ${\rm d}r/{\rm d}z$ scales as
$\hinv$, $M\,n(M)$ scales as $h^3$, and $\sdla$ scales as $h^{-2}$ in
the simulation.  For the cumulative version of this calculation, see
Equation~(8) and Figure~5 of \citet{NSH04a}. Equation~(\ref{eq:rate})
can be derived from the following expression for the DLA area covering
fraction on the sky along the line element $c\,\dd t$: \beq \dd N_{\rm
DLA} &=& n_{\rm phys}(M)\,\dd M \cdot \sdlaphys\,\cdot c\, \dd t
\label{eq:cover}\\ &=& (1+z)^3\, n_{\rm co}(M)\,dM \cdot \sdlaphys
\cdot a\,\dd r \\ &=& n_{\rm co}(M)\,dM \cdot \sdlaco \cdot \dd r,
\eeq where we have used $c\,\dd t = a\,\dd r$ and $\sdlaco =
(1+z)^2\,\sdla^{\rm phys}$. Here $a$ is the scale factor, $\dd r$ is
the line element in comoving coordinate, and $n_{\rm co}(M)\,{\rm d}M$
and $n_{\rm phys}(M)\,\dd M$ are the comoving and physical number
density of halos in the mass range [$M$, $M+{\rm d}M$], respectively.
Sometimes the `absorption distance' ${\rm d}X$ is defined as ${\rm d}X
\equiv \frac{H_0}{c}\,(1+z)^3\, c\,\dd t = \frac{H_0}{c}\,(1+z)^2\,
\dd r = \frac{H_0}{H(z)}\,(1+z)^2\, {\rm d}z = (1+z)^2\, \dd z / E(z)$, and
is used to express the rate-of-incidence as \beq \frac{\dd N_{\rm
DLA}}{{\rm d}X} = \frac{c}{H_0}\, n_{\rm co}(M)\,{\rm d}M \cdot \sdla^{\rm phys}.
\label{eq:dX}
\eeq
For $z=3$ and our adopted cosmology, ${\rm d}X/{\rm d}z = 3.5867$. 
In Equations~(\ref{eq:cover}) to (\ref{eq:dX}), we left in the dependence
on halo masses explicitly, but in practice an integral over a 
certain range of halo mass has to be performed when comparing 
with actual observations. 

We now use the power-law fits for $\sdlaco(M,z)$ described above 
to compute the differential distribution of DLA incidence 
via Equation~(\ref{eq:rate}). The result is shown in 
Figure~\ref{fig:pdf} for all the simulations at $z=3$. The qualitative 
features of the curves are easy to understand. Because 
$n(M)\propto M^{-2}$ at $M\approx 10^8 - 10^{12}\himsun$ 
(see Figure~\ref{fig:st}), the 
distribution is flat when $\sdla \propto M$. In fact, $n(M)$ is 
slightly shallower than $M^{-2}$ (more like $M^{-1.8}$), therefore 
the distribution for the P3 run is almost flat at $10^8< M < 
10^{12}\himsun$,  because $\sdla \propto M^{0.79}$ in this
simulation.  At masses higher than $10^{12}\himsun$, 
the mass function deviates from the $M^{-2}$ power-law significantly, 
and the distributions for all runs quickly drop off to a small value. 

The halo masses where each distribution peaks are listed in the second
column of Table~\ref{table:meanmass}. The peak halo mass $\Mpeak$ 
becomes larger as the feedback strength increases. For the O3 run, 
we indicated $\Mpeak = 10^{8.5}\himsun$ in parentheses because
we think that the DLA cross section rapidly falls off at this 
halo mass based on the work by \citet{NSH04a} and the peak halo mass 
is simply this cutoff mass-scale.  
The peak halo mass is significantly larger for the Q4 
($\Mpeak = 10^{11.6}\himsun$) and Q5 ($\Mpeak = 10^{12}\himsun$)
runs compared to other runs. 


\section{Mean \& Median halo masses of DLAs} 

\begin{figure}
\begin{center}
\resizebox{14cm}{!}{\includegraphics{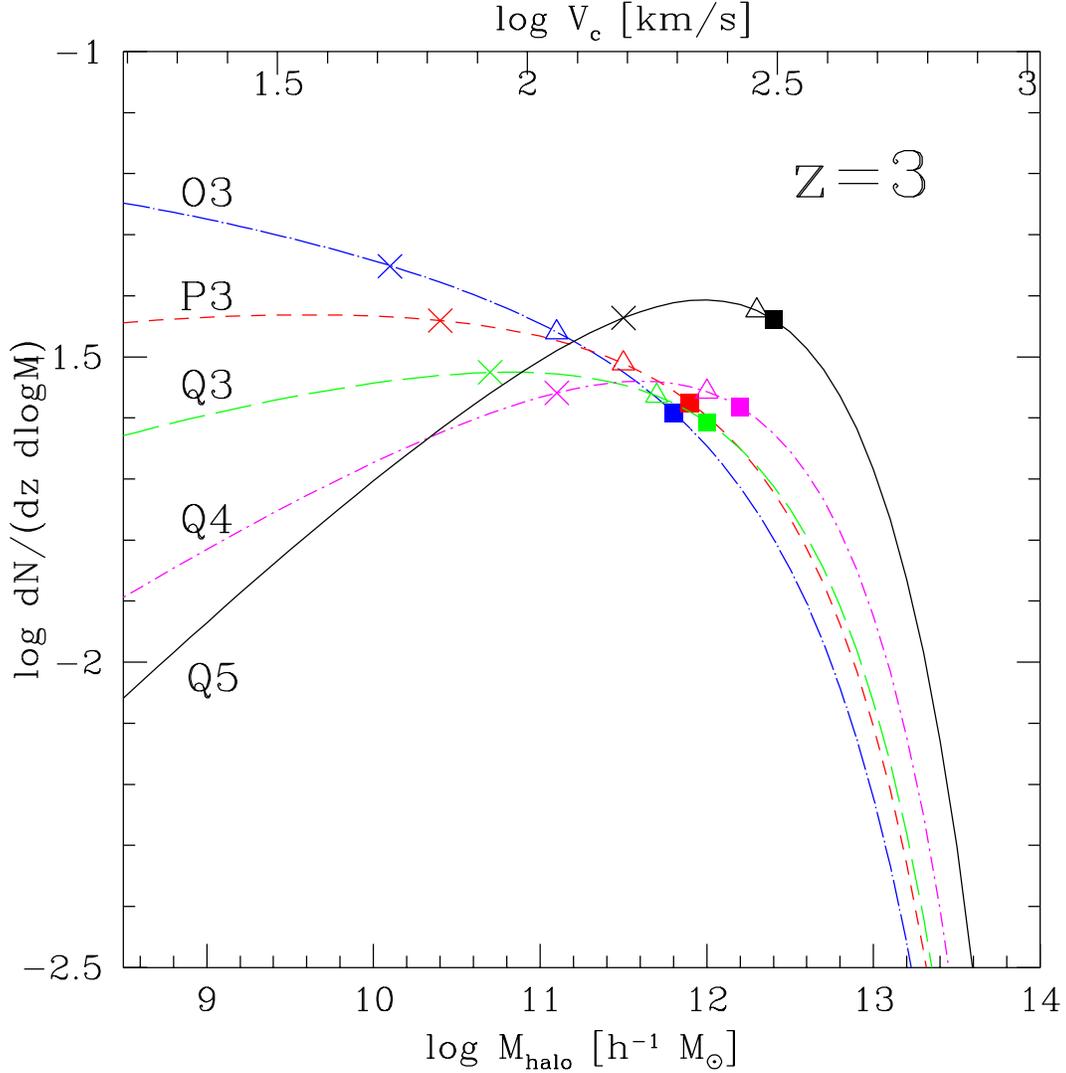}}
\caption{Differential distribution of DLA rate-of-incidence 
$dN/(dz\,d\log M)$ at $z=3$ as a function of halo mass, computed
using Equations~(\ref{eq:sigma}) and (\ref{eq:rate}).
The open crosses indicate the median halo masses $\Mmed$,  
and the open triangles indicate the 75 percentile of the 
distribution $\Msev$. 
The filled squares indicate the mean DLA halo masses $\Mmean$. 
The top axis also indicates the circular velocity as computed by 
Equation~(\ref{eq:vc}).
The differences in the shape of the distribution arise from
the differences in the relationships between DLA cross section
and halo mass as given by Equation~(\ref{eq:sigma}).  
}
\label{fig:pdf}
\end{center}
\end{figure}

For each distribution shown in Figure~\ref{fig:pdf}, 
we compute the mean DLA halo mass as
\beq
\quad \Mmean &=& 
\frac{\int_0^{\infty}M\,\frac{\dd N}{\dd z \dd \log M} \dd \log M}{\int_0^{\infty}\frac{\dd N}{\dd z \dd \log M} \dd \log M} \\
&=& \frac{\int_0^{\infty}M^2\, n(M)~\sdla(M)~\dd\log M}{\int_0^{\infty}M\, n(M)~\sdla(M)~\dd\log M}, 
\label{eq:mean}
\eeq
and the result is summarized in Table~\ref{table:meanmass}.
The mean halo mass is smaller for the `no-wind' (O3) run, 
and is larger for the `strong-wind' (Q3 to Q5) runs. 
This is because of the steepening of the relationship between 
$\sdla$ and $\Mhalo$ as the feedback strength increases. 
But the mean halo mass is in the range $\Mdla = 10^{11.5} - 
10^{12.5}\himsun$ for all runs.  While this mass-scale is not 
as large as a Milky Way sized halo 
($\Mhalo \simeq 10^{12}-10^{13}\Msun$), it is certainly more massive 
than that of dwarf galaxies. We also list the mean of $\log \Mdla$ 
for comparison with the results by \citet{Bouche05}. We will discuss 
the implications of these mass-scales in Section~\ref{sec:discussion}.
 
One can also look at the median halo mass $\Mmed$, 
below (or above) which 50 percent of the DLA incidence is covered. 
The quantity $\Mmed$ is defined by the following equation:
\beq
0.5 = \int_{\Mmed}^{\infty}n(M)~\sdla(M)~\dd M ~/~ 
\int_0^{\infty}n(M)~\sdla(M)~\dd M. 
\eeq
The median halo mass is always smaller than $\Mmean$ for all runs. 
The values of $\Mmed$ are indicated by the open crosses in 
Figure~\ref{fig:pdf}, and $\Mmean$ are indicated by the filled squares. 
We emphasize that $\Mmed$ is much smaller than $\Mmean$, and that 
$\Mmean$ could be biased towards a large value because of the 
weighting by the halo mass.

Furthermore, in Table~\ref{table:meanmass} we also give the 75 
percentile halo mass $\Msev$, below which 75 percent of the DLA 
incidence is covered. For example, in the Q5 run, halos with masses 
$\Mhalo < 10^{12.3}\himsun$ are responsible for 75\% of the DLA
incidence at $z=3$. The values of $\Msev$ are shown with open 
triangles in Figure~\ref{fig:pdf}.


\section{Luminosity distribution of DLA galaxies}

\begin{figure}
\begin{center}
\resizebox{14cm}{!}{\includegraphics{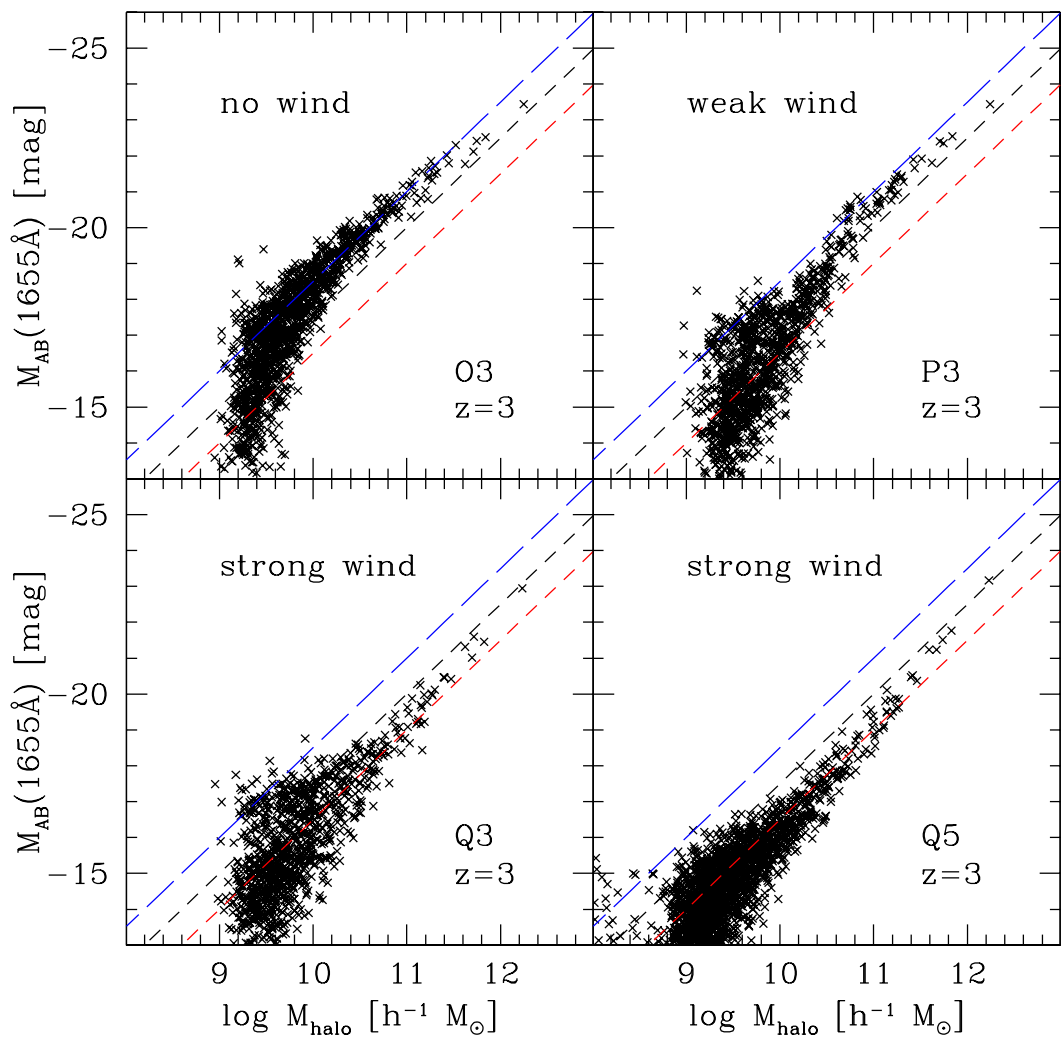}}
\caption{Halo mass vs. absolute AB magnitude at 1655\AA\, at $z=3$. 
See text for the description of the three dashed lines. }
\label{fig:mab}
\end{center}
\end{figure}

The luminosity distribution of DLA galaxies constrains their nature, 
facilitating the interpretation of observations of DLA galaxies. 
\citet{NSHM} computed the spectra of galaxies 
in the same simulations used in this paper using 
the population synthesis model of \citet[][hereafter BClib03]{BClib03} 
based on the stellar mass, formation time, and metallicity 
of each stellar particle that makes up simulated galaxies. 
Using the computed spectra,
we sum up the monochromatic luminosity at rest-frame 1655\AA\, 
(chosen because it corresponds to the observed-frame ${\cal R}$ band
of the $U_nG{\cal R}$ system) 
of all the galaxies enclosed in the maximum radius 
of each dark matter halo identified by the 
friends-of-friends grouping algorithm \citep{Davis85}.  
The following formula \citep[see Eq.(2) of][]{Night06}
is used  to compute the absolute AB magnitude at 1655\AA, 
because BClib03 outputs its spectrum $L_\lam$ in units of 
$\Lsun$\,\AA$^{-1}$ where $\Lsun=3.826\times 10^{33}$\,erg\,s$^{-1}$: 
\beq
M_{AB} = -2.5\log (\lam^2 L_\lam) + 13.83. 
\label{eq:mabsolute}
\eeq
Figure~\ref{fig:mab} shows the relationship between $M_{AB}$ and 
the halo mass for O3, P3, Q3, and Q5 runs at $z=3$. 
The three dashed lines correspond to 
\begin{equation}
M_{AB} = -2.5\, (\log \Mhalo - 12) - C_1,
\label{eq:mab}
\end{equation}
where $C_1 = 23.5, 22.5$, and 21.5 from top to bottom, and
$\Mhalo$ is in units of $\himsun$. 
We adopt $C_1=23.5$ for the O3 run, $C_1=22.5$ for the P3 run, 
and $C_1=21.5$ for the Q3 and Q5 runs. 
This figure shows that, on average, the galaxies in the `no-wind' run (O3) 
are brighter than those in the `strong-wind' run (Q3 \& Q5) 
by about a factor of six. This was pointed out in Figure~6 of \citet{NSHM}
and was attributed to the suppression of star formation by a
strong galactic wind. 

Here, the apparent and absolute magnitudes are related to each 
other as
\beq
m_{AB} = M_{AB}+ 2.5\, \log(1+z) + 5\,\log(d_L/10{\rm pc}),
\label{eq:mapparent}
\eeq 
where $d_L$ is the luminosity distance. Notice the positive sign 
in front of the 2nd term (which is normally negative); this 
is because of the definition of $M_{AB}$ in Equation~(\ref{eq:mabsolute})
\citep[see also Equation~(2) of][]{Night06}. 
Inserting Equation~(\ref{eq:mab}) into Equation~(\ref{eq:mapparent})
and using $d_L(z=3) = 2.542\times 10^4\, h_{70}^{-1}\,\mpc$ for our flat 
$\Lam$ cosmology, 
we obtain the relation between the apparent magnitude $\Rab$ and
the halo mass $\Mhalo$ as 
\begin{equation}
\Rab = -2.5\,\log \Mhalo + C_2 - 5\log h_{70},
\label{eq:Rab}
\end{equation}
where $C_2=55.03$ (O3 run), 56.03 (P3 run), and 57.03 (Q3 and Q5 run), 
and $\Mhalo$ is in units of $\himsun$.  
These numerical values are consistent with the ones adopted 
by \citet{Hae00}, where they assumed $m_{AB}=26.6 - 7.5 \log(v_c/200\kms) 
= -2.5\,\log \Mhalo + 55.7$ for a flat $\Lam$ cosmology. 
The circular velocity $v_c$ at a radius of overdensity 200 is 
computed as 
\begin{eqnarray}
v_c &\equiv& \left( \frac{G\Mhalo}{R_{200}}\right)^{1/2}
= \left[ G \Mhalo^{2/3} \left(\frac{4\pi}{3}\bar{\rho}\,200\right)^{1/3} \right]^{1/2} \\ 
&=& 123.5\, \left( \frac{\Mhalo}{10^{11} \himsun}\right)^{1/3} \left(\frac{1+z}{4}\right)^{1/2} \kms 
\label{eq:vc}
\end{eqnarray}
for our flat $\Lam$ cosmology and $\bar{\rho}$ is the mean density
of the universe at redshift $z$.

Figure~\ref{fig:cum_abs} shows the cumulative distribution for the
absolute value of DLA rate-of-incidence as a function of apparent
$\Rab$ magnitude of host galaxies.  The slight differences in these
results from Fig. 5 of \citet{NSH04a} are due to an improved
calculation of the halo mass function with a more accurate integration
of the power spectrum when computing the mass variance $\sigma(M)$.
The yellow hatched region shows the observational estimate $\log ({\rm
d}N/{\rm d}z) = -0.60 \pm 0.10$ at $z=3$ obtained from Fig.~8 of
\citet{Pro05}.  We estimate that the range of ${\rm d}N/{\rm d}X$
centered at $z=3$ is [$0.065, 0.09$] with a central value $0.07$, and
then multiply ${\rm d}N/{\rm d}X$ by ${\rm d}X/{\rm d}z = 3.587$ for
$z=3$ to obtain the above observational value.  We find that the P3,
Q3 and Q5 runs underpredict ${\rm d}N/{\rm d}z$. We will discuss this discrepancy
further in Section~\ref{sec:discussion}.

It is clear from Figure~\ref{fig:cum_abs} that only a very small
fraction of DLA incidence is associated with galaxies brighter than
$\Rab = 25.5$.  This is more evident in Figure~\ref{fig:cum_norm}
where the cumulative probability distribution of DLA incidence is
shown.  In panel (a), we normalize the cumulative distribution by the
value at $\Mhalo = 10^{9.8}\himsun$, or equivalently, halos less
massive than the above value are assumed not to host DLAs at all.
Here we roughly reproduce the result of \citet{Hae00} where they
adopted the cutoff circular velocity of $50 \kms$, which corresponds
to $\Mhalo \simeq 10^{9.8}\himsun$.  Similarly to their result, we
find that for this cutoff mass, only $10-20$\% of DLA sight-lines are
contributed by galaxies brighter than the spectroscopic limit $\Rab =
25.5$, and $70-90$\% of the DLA sight-lines are contributed by
galaxies brighter than magnitude $\Rab = 30$.

However, according to the work by \citet{NSH04a}, 
only halos with masses $\Mhalo \gtrsim 10^{8.5} \himsun$ 
(or equivalently $v_c > 18\kms$) contribute to 
the DLA cross section (see their Figure 2 \& 3), 
so the cutoff velocity of $50\kms$ adopted by \citet{Hae00} 
might be too high.  In Figure~\ref{fig:cum_norm}b 
we show a case with cutoff halo mass $\Mhalo >10^{8.5}\himsun$. 
Here, less than 15\% of DLAs are contributed by galaxies 
with $\Rab<25.5$, and $50-70$\% by those with $\Rab<30$.


\section{Impact parameter distribution}
\label{sec:impact}

We also compute the impact parameter distribution (i.e., projected
separation between DLA sight-lines and the nearest galaxy) using our
simulations.  \citet{NSH04a, NSH04b} calculated the neutral hydrogen
column density $\NHI$ for each pixel with a size of $\epsilon^2$ on
projected planes covering the face of each dark matter halo, where
$\epsilon$ is the smoothing length of the simulation.  The pixels with
$\NHI > 2\times 10^{20}$ cm$^{-2}$ were equally regarded as DLA
sight-lines.  Therefore there are multiple DLA sight-lines per halo.
Galaxies in our simulations are identified as collections of star
particles.  \citet{NSHM, NSHM2} computed the spectrophotometric
properties of simulated galaxies based on the formation time, stellar
mass, and metallicity of individual star particles, and showed that
the simulated luminosity functions agree reasonably well with the
observations if a mean extinction of $E(B-V)=0.15$ is assumed.  This
level of extinction is consistent with observationally derived values
for the Lyman break galaxies (LBGs) at $z\sim 3$ \citep{Shapley01}.

Knowing the locations of both DLAs and galaxies in the simulation, 
we compute the impact parameter for each DLA pixel by searching 
for the nearest galaxy on the projected plane. 
If a nearby galaxy cannot be found within the same halo, 
we allow the search to extend further.
Figure~\ref{fig:impact_norm} shows the cumulative probability
distribution of DLA incidence as a function of impact parameter 
$\bphys$ (in units of physical $\hsevkpc$). 
Figure~\ref{fig:impact_norm}a shows the results of the O3, P3, and 
Q3 runs to highlight the impact of galactic wind feedback. 
As the feedback strength increases, gas in low mass halos is
ejected more efficiently, and the neutral hydrogen content decreases. 
Therefore, the relative contribution from higher mass halos
increases and the impact parameter distribution becomes broader.
Another notable feature of this plot is that, if all galaxies
are allowed to be a candidate DLA galaxy no matter how faint they are, 
then the majority (over 90\% for the O3 run, and 80\% for the P3 
and Q3 runs) of DLA sight-lines have the nearest galaxy within 
$\bphys = 5$\,$\hsevkpc$. 
However, if we limit the search for the nearest galaxy to those brighter 
than $\Rab=30$ or 28 mag, then a large fraction of DLAs, in 
particular those in low mass halos, cannot be associated with
a qualified galaxy within the same halo, resulting in a much broader 
impact parameter distribution.  
About 30\% of all DLA sight-lines in the O3 run have 
impact parameters $\bphys > 5$\,$\hsevkpc$ for the limited search 
of a nearest galaxy with $\Rab < 28$. 

Figure~\ref{fig:impact_norm}b shows the result of the Q3 and Q5 runs. 
The higher resolution run (Q5) has a narrower impact parameter
distribution than the lower resolution run (Q3), because
it can resolve more low-mass galaxies which host DLAs with 
low impact parameters.  Therefore the relative contribution from 
DLAs with low impact parameters becomes larger. Also, galaxies
in massive halos will be better resolved in the Q5 run than in  
the Q3 run, and this will also result in smaller impact parameters
for the DLAs in massive halos.
We expect that, in a run with even higher resolution,
the impact parameter distribution will remain as narrow as 
that of the Q5 run, 
because increasing the resolution always seems to increase 
the relative number of columns with low impact parameters. 
We also show the case where we limit the search for the nearest 
galaxy to those brighter than $\Rab=30$ mag. 
In this case, results from the Q3 and Q5 runs agree well, and  
about 40\% of all DLA sight-lines have impact parameters 
$\bphys > 5$\,$\hsevkpc$. 

For most of the DLAs with low impact parameters, there is a 
galaxy within the same dark matter halo. However, for a few to 
10\% of the DLA sight-lines that are in the low mass halos 
in O3, P3, and Q3 runs, there are no galaxies within the same halo 
and the nearest galaxy is in another halo, as indicated by the 
offset of the curves at the bottom right corner of the plot in 
panel (a). In the Q5 run, there are almost no DLA sight-lines 
that do not have a galaxy within the same halo, and 
the distribution approaches zero at large $\bphys$ values 
in panel (b). 

Overall, our impact parameter distribution seems to be much narrower
than that obtained by \citet{Hae00} if we do not limit our search 
to the bright galaxies.  Figure~3 of \citet{Hae00} suggests that 60\% 
of DLA sight-lines have $b<1$ arc second, but our results indicate 
that more than 80\% of DLAs have $b<1$\,arc second if we do not 
limit our search for the nearest galaxy to those with $\Rab<30$ mag. 
The differences between the two results might stem from subhalos 
within the massive halos in the simulations,  which \citet{Hae00} 
did not take into account.  The calculation of \citet{Hae00} assigns 
a single effective DLA radius to each halo and assumes that the DLAs 
cover a circular area centered on each halo, whereas in our simulations, 
massive halos contain numerous galaxies and the geometry of the 
DLA cross section cannot be characterized by a circular area 
centered on each halo. For example, the most massive halo in 
the O3 run contains 143 galaxies, and in the Q5 run 1110 galaxies. 
Therefore the DLAs in the simulation would be able to 
find the nearest galaxy at distances much smaller than the effective 
DLA radius computed by \citet{Hae00}. But when we limit our search
for the nearest galaxy to those with $\Rab<30$ mag, then our results
for the Q3 and Q5 run become similar to that obtained by \citet{Hae00}.
It is not very clear how our results compare to Figure~11d by \citet{Mo98}, 
as these authors examined only the differential distribution
of impact parameter and not the cumulative probability distribution.
But their differential distribution peaks at 2.5 $h^{-1}$\,kpc and a
significant fraction (more than 50\%) of the area under the curve
appears to come from $b<5\,h^{-1}$\,kpc, in reasonable agreement 
with our results.


\section{Number density of DLAs}
\label{sec:ndla}

Finally, we discuss the comoving number density of DLAs. Assuming that 
the characteristic covering area of each DLA $\Adla$ is fixed with 
a physical radius $\rdla$ (i.e., $\Adla = \pi\,\rdla^2$), we can 
compute the cumulative number density of DLAs as a function of halo 
mass as
\beq
N_{\rm DLA}(>M) = \int_M^{\infty} {\rm d}M\, n(M)\, \frac{\sdlaphys}{\Adla},
\eeq
where the definitions of $n(M)$ and $\sdlaphys$ are the same as described
in Section~\ref{sec:rate}.  Note again that $\sdla$ is the {\it total}
DLA cross section of each dark matter halo; in other words, if there
are 100 DLA gas clouds in a massive halo, then this halo has a total DLA cross
section of $\sdlaphys = 100\, \Adla$. 
Then, using Equation~(\ref{eq:mab}) allows us to obtain the cumulative 
comoving number density of DLAs as a function of 
apparent $\Rab$ magnitude $N_{\rm DLA}(<\Rab)$ as shown in 
Figure~\ref{fig:ndla}. Here, three different values of physical radius
for DLAs are assumed: $\rdla = 1, 5$ and $20\,\hsevkpc$. For each 
radius, the results of four simulations (O3, P3, Q3, and Q5 runs) are 
shown. Also given is the cumulative number density of LBGs, 
$N_{\rm LBG}(<m)$, computed by integrating the Schechter luminosity 
function of \citet{Ade00} with parameters $(m^*, \alpha, 
\Phi^*[h^3\mpc^{-3}]) = (24.54, -1.57, 4.4\times 10^{-3})$:
\beq
N_{\rm LBG}(<m) = \int_{m}^{0}\Phi(m){\rm d}m, 
\eeq
where
$\Phi(m) = (0.4\,\ln 10)\,\Phi^*\,10^{\mu\,(\alpha + 1)}\,\exp(-10^\mu)$
as a function of apparent magnitude $m$, and $\mu = 0.4\,(m^* - m)$. 
The comoving number density of LBGs at $z=3$, $N_{\rm LBG} = 4\times 10^{-3}$ 
\citep{Ade03}, is also shown as the data point at the magnitude limit 
of $\Rab = 25.5$.

Figure~\ref{fig:ndla} shows that the comoving number density of DLAs
is larger than that of LBGs down to the magnitude limit of $\Rab=25.5$
if the physical radius of each DLA is smaller than $\rdla \simeq
5\,\hsevkpc$. The two number densities roughly agree with each other
at $\Rab = 25.5$ when $\rdla \simeq 20\,\hsevkpc$.  Earlier,
\citet{Schaye01} argued that the observed DLA rate-of-incidence can be
accounted for if each LBG were accompanied by a DLA cross section of
$\pi r^2 = \pi (19\hikpc)^2$ assuming $({\rm d}N/{\rm d}z)_{\rm
DLA}=0.20$ and $N_{\rm LBG} = 0.016\, h^3\,\mpc^{-3}$ (down to
$\Rab=27$ mag).  His result is in good agreement with the result of
the Q5 run with $\rdla = 20\,\hsevkpc$ shown in Figure~\ref{fig:ndla}.
However, this large DLA radius is somewhat unrealistic because this
model implies that all halos with masses $\Mhalo \gtrsim
10^{12}\himsun$ host such a large disk at $z=3$ that are responsible
for DLAs, and none of the less massive halos host DLAs at all. This
picture is quite the contrary to our simulation results that indicate
halos down to masses $\Mhalo = 10^{8.5}\himsun$ could host DLAs, and
simulated DLAs in lower mass halos are clumpy and smaller in cross
section.

The existence of extended disks at high redshift can be observationally 
tested by searching for extended emission from stars in deep imaging 
data such as the Hubble Deep Fields \citep{Bouwens03, Ferguson04}. 
The most recent study on this issue by \citet{Bouwens04b} using the
Hubble Ultra Deep Field (HUDF) suggests that high redshift $UBVi$-dropout 
galaxies are compact in size ($\sim 0.1 - 0.3$ arc seconds) and that 
extended sources ($\gtsim 0.4$ arc sec, $\gtsim 3\,\kpc$) are rare. 
Another observational analysis of high redshift galaxies in HUDF by 
\citet{Wolfe06} also suggests that there are 
no extended disks down to very faint surface brightness. 
Our simulation results and impact parameter distribution are in accord 
with these observational results. If our picture is realistic and each DLA 
has a physical size of $\ltsim 5\,\kpc$, then it means that there
are multiple clumps of DLAs around massive galaxies such as LBGs
at $z\sim 3$, although the fraction of DLA incidence covered by 
the DLAs in massive halos ($\Mhalo \gtsim 10^{12}\himsun$) is 
smaller than that in less massive halos that host very faint 
galaxies ($\Rab \gtsim 27$\,mag).


\section{Discussion \& Conclusions}
\label{sec:discussion}

Using state-of-the-art cosmological SPH simulations, 
we performed a numerical study of a galactic wind model, and 
examined the distribution of DLA rate-of-incidence as a function of
halo mass, galaxy apparent magnitude, and impact parameter. 
We find that the majority of DLA rate-of-incidence in the 
simulations is dominated by relatively lower mass halos 
($\Mhalo < 10^{12}\himsun$) and faint ($\Rab>25.5$) galaxies. 
This conclusion agrees with the generic prediction 
of the semi-analytic model of \citet{Kau96}, that the DLA galaxies
at high redshift will typically be smaller, more compact, and less
luminous than disk galaxies at the present epoch, although this 
analysis was restricted to an $\Om=1$ universe. 
More recent work by \citet{Okoshi05} also suggests that the low-redshift
($z\le 1$) DLA galaxies are mainly low-luminosity, compact galaxies. 
Combined with our results, the dominance of faint galaxies in DLA 
incidence seems to be a generic prediction of a CDM model. 
This can be ascribed to the steeply rising dark matter halo mass function 
towards lower masses in CDM models, and to the fact that the small 
DLA cross sections in these low mass halos add up to a large portion
of the total DLA incidence when multiplied by a large number of 
low-mass halos. 


\subsection{On the DLA Halo Mass}

We characterize the differential distribution of DLA rate-of-incidence
$dN/(dz\,d\log M)$ with various halo masses listed in 
Table~\ref{table:meanmass}. 
We find that the mean DLA halo mass increases with increasing galactic wind 
feedback strength, because winds are able to eject the gas in 
lower mass halos, suppressing their DLA cross section,
resulting in a larger relative contribution from higher mass halos 
\citep[see also][]{NSH04a}. 
The mean DLA halo mass for the Q5 run was found to be 
$\Mmean = 10^{12.4}\himsun$ and 
$\avg{\log M_{\rm DLA}[\himsun]} = 11.3$ when 
we limit the DLA distribution to $\Mhalo > 10^{8.5}\himsun$.  
The latter value is close to that obtained by 
\citet[][$\avg{\log M_{\rm DLA}[\himsun]} = 10.9$]{Bouche05}, 
but this comparison is not fully appropriate because their simulation 
only resolved halos with masses  
$\log (\Mhalo [\himsun])> 10.5$ and they did not attempt to 
extrapolate their halo distribution using an analytic halo mass function.
Therefore, the value of mean DLA halo mass they derived 
was biased toward a larger value owing to limited resolution, and 
it is an upper limit as the authors described in their paper. 
Since their simulation did not include galactic wind feedback, 
their result should be compared to that of our O3 (`no-wind') run. 
In fact, the value of concern for the O3 run is 
$\avg{\log M_{\rm DLA}[\himsun]} = 10.4$, and it is 
lower than their halo mass resolution when we take the full 
distribution of DLAs into account down to $\Mhalo = 10^{8.5}\himsun$
We also note that there are large differences between 
$\lgMmean$, $\log \Mmean$, \& $\Mmed$, therefore the former two 
quantities are not to be confused with $\Mmed$.

Given that the results of the Q3 and Q5 runs do not agree, 
the results given here may appear as if they have not converged yet. 
According to the results of the R-Series presented by \citet{NSH04a}, 
we consider that the lowest halo mass that could host DLAs is  
$\Mhalo \sim 10^{8.5}\himsun$. 
In the Q5 run, there are 150 dark matter particles for a halo of this mass, 
therefore it is close to the resolution limit. 
The Q5 run should be close to the convergence if not fully converged. 
It is possible that in a run with higher resolution run than the Q5 run, 
the mean DLA halo mass could be even higher than that of the Q5 run. 

There have already been several observational attempts to constrain 
DLA halo masses via cross-correlation between DLAs and LBGs
\citep{Gawiser01, Ade03, Bouche03, Bouche04}, but owing to limited sample 
sizes, the results have been mostly inconclusive. More recently, 
\citet{Cooke06} has measured the cross-correlation between 11 DLAs 
and 211 LBGs, and constrained the DLA halo mass to be 
$\approx 10^{11.2}\Msun$. It is encouraging that this measurement is 
close to our results listed in Table~\ref{table:meanmass}.  
The result of \citet{Cooke06} suggests that at least some DLAs 
are associated with relatively massive halos, 
close to the LBG halo masses ($\sim 10^{12}\Msun$). 
There could of course be some distribution in the LBG halo masses 
(perhaps from $\Mhalo=10^{10}$ to $10^{13}\Msun$; see the broad 
distribution of stellar masses at the magnitude limit $\Rab = 25.5$ 
in Figure~4 of \citet{NSHM}), and likewise DLAs could also have 
a broad halo mass distribution. 
The majority of DLA sight-lines in the simulations are 
dominated by lower mass halos in spite of a relatively large 
mean DLA halo mass, which is also reflected in the large 
differences between the median halo mass and the mean halo 
mass as we summarize in Table~\ref{table:meanmass}.


\subsection{On the Luminosity Distribution}

As for the luminosity distribution of DLA galaxies, we find that only
about 10\% of DLA sight-lines are associated with galaxies brighter
than $\Rab = 25.5$ mag. This suggests that only about 10\% of
DLA galaxies will be found in searches for the bright LBGs at
$z\sim 3$ down to $\Rab=25.5$. The dominance of DLA sight-lines by 
the faint galaxies is a generic result in all of our simulations, 
therefore we consider that this conclusion will not change 
in a higher resolution run than the Q5 run.  

We reproduce the result of \citet{Hae00} when we cut off the 
DLA distribution at $\Mhalo = 10^{9.8}\himsun$ (or equivalently a 
circular velocity of $v_c = 50\,\kms$), and in this case $70-90$\% of 
DLA sight-lines are associated with galaxies brighter than $\Rab=30$ mag. 
This agreement with the Haehnelt et al. result is not surprising, 
because they assumed a relation $\sdla \propto v_c^{\alpha} \propto
\Mhalo^{\alpha/3}$ with $\alpha \sim 2.5-3$ and a normalization
matched to the observed $dN/dz$ of DLAs, and our SPH simulations
suggest $\alpha = 2.2 - 3.1$ depending on the feedback strength and
resolution.  
However, our simulations indicate that even lower mass
halos contribute to the DLA cross section, and when we cut off our DLA
distribution at $\Mhalo = 10^{8.5}\himsun$, only $50-70$\% of DLAs are
associated with galaxies brighter than $\Rab = 30$ mag.
Thus, it would be possible to detect DLA galaxies with $\sim 50$\%
efficiency by searching down to $\Rab=30$ mag.

While the dominance of faint galaxies among DLAs
seems to be a generic prediction of the CDM model, we note that
the recent photometric survey of low-redshift ($z<1$) DLA galaxies 
by \citet{Chen03b} on the contrary suggests that a large contribution 
from dwarf galaxies is not necessary to account for their observed
DLA incidence. However, this conclusion might 
be affected by the `masking effect' emphasized by \citet{Okoshi05}.  
Current DLA surveys might be biased against the 
detection of DLAs associated with faint and compact galaxies, 
because such galaxies would be buried under the bright QSO 
that has to be masked for the detection of a DLA galaxy. 
This effect would be more severe if the impact parameters 
are small as our present work suggests, as well as 
some observational studies of high redshift DLA galaxies 
that imply very small impact parameters less than several kpc (i.e., $\lesssim 1$ 
arcsec) \citep{Fynbo99, Kulkarni00, Moller02, Moller04}. 
\citet{Fynbo99} suggests that, based on the properties of 
a limited sample of observed high redshift candidate DLA galaxies, 
a large fraction ($\sim 70$\%) of DLA galaxies at $z\simeq 3$ 
could be fainter than $\Rab=28$ mag. 
It is possible that evolution with redshift is relatively strong
as suggested by the chemical evolution model of \citet{Lanf03}, 
in the sense that the high redshift DLAs are dominated by 
dwarf galaxies and the low-redshift ones by disks. 
But the latter scenario seems to be inconsistent with 
the predictions by \citet{Okoshi05}. 

\citet{Hopkins05} also argued for the dominance of faint galaxies
for the DLA galaxies based on the comparison of global quantities such 
as the density of gas mass, stellar mass, metal mass, and star 
formation rate. They also suggested that the DLAs may be a distinct
population from LBGs, but we note that the dominance of faint galaxies
for the DLA rate-of-incidence does not immediately mean that DLAs 
do not exist in LBGs. It simply means that the area covered by the 
DLAs associated with LBGs is much smaller than that in faint galaxies. 
We plan to investigate the connection between DLAs and LBGs in more 
detail in future work.


\subsection{On the Impact Parameter Distribution}

Our distribution of impact parameters is 
significantly narrower than that of \citet{Hae00}, and we ascribe
this difference to substructures within the massive halos
in our simulations.
The differential distribution of 
\citet[][Figure~11d]{Mo98} seems to be in better agreement with our 
results, with roughly half of DLAs having $\bphys < 3\hikpc$, but this 
comparison is probably not fully appropriate because their model is 
restricted to centrifugally supported disks. 
Indeed, \citet{Hae00} comment that their effective DLA radius 
is about a factor of 10 larger than the 
expected scale length of a centrifugally supported disk 
if the angular momentum of the gas owes to tidal torquing 
during the collapse of a dark matter halo. 
However, we caution that the large effective DLA radius for a massive 
halo does not necessarily mean that DLAs consist of only large 
disks centered on halos; i.e., the DLA cross section could be
distributed to hundreds of galaxies (associated with subhalos) 
embedded in a massive halo, with each DLA gas cloud being fairly compact. 
This point was also emphasized by \citet[][Section~4.1]{Gardner01}. 

Our narrow impact parameter distribution at first sight might seem 
inconsistent with that of \citet{Gardner01}, but in fact they are not. 
\citet{Gardner01} mostly used SPH simulations with 
64$^3$ particles which were able to resolve halos only down to 
$\Mhalo = 8.2\times 10^{10}\Msun$, and found that nearly all DLAs 
lie within $15-20$ kpc of a galaxy center. Given their mass resolution, 
they were not able to simulate galaxies fainter than $\Rab \simeq 30$ mag
at $z=3$. If we limit the search for the nearest galaxy to those brighter 
than $\Rab = 30$ mag (see Figure~\ref{fig:impact_norm}b), our simulations 
suggest that 80\% of DLAs are within $\bphys = 15\,\hsevkpc$, 
which is consistent with the results of \citet{Gardner01}. 
However if we include the fainter galaxies that were not resolved
in the simulations by \citet{Gardner01}, then the overall 
impact parameter distribution becomes narrow, 
with more than 80\% of DLA sight-lines having $\bphys < 5\,\hsevkpc$. 

The small impact parameters suggest that DLAs in our simulations are
compact, and this has significant implications for the nature of
DLAs. The compactness of the simulated DLAs can also be observed in
the projected distribution of DLAs in Figure~2 \& 3 of \citet{NSH04b}.
There, one can see that the DLAs are at the centers of galaxies and
coincide with star-forming regions quite well, covering roughly half
the area of the star-forming region. These results are at odds with
the idea that DLAs mainly originate from gas in large galactic disks
\citep[e.g.,][]{Wolfe86, Turnshek89, Pro97, Pro98}.  The overall
agreement between the simulations and observations found by
\citet{NSH04a,NSH04b} was encouraging, but the simulations with
galactic wind feedback now seem to underpredict the DLA
rate-of-incidence (see Figure~\ref{fig:cum_abs}) compared to the new
observational estimate by \citet{Pro05} utilizing the SDSS Data
Release 3.  This is because our simulations underpredict the column
density distribution $f(\NHI)$ at $20 < \log \NHI < 21$ by a factor of
$2-3$ even in our highest resolution run with a smoothing length of
comoving 1.23\,$\hikpc$.  \citet{Cen03} reported that they did not
have this problem, but instead they significantly overpredicted
$f(\NHI)$ at $\log \NHI > 22$ as well as $\Omega_{\rm gas}$, and had
to introduce dust extinction to match observations.  The inconsistency
in $f(\NHI)$ between SPH simulations and observations and the
compactness of the simulated DLAs might be related to the well-known
angular momentum problem \citep[][and references therein]{Robertson04}
in hydrodynamic simulations, where simulated disk galaxies are known
to be too small and centrally concentrated.  This is a problem that is
not easy to solve, and future studies are needed using both Eulerian
mesh simulations and SPH simulations.  In summary, there persist
tensions between the predicted and observed $f(\NHI)$, large observed
velocity widths (which favors large and thick disks), and the
predicted compactness of DLAs \citep[cf.][]{Jedamzik98}.


\subsection{On the Resolution Effects and Numerical Convergence}

\begin{figure}
\begin{center}
\resizebox{14cm}{!}{\includegraphics{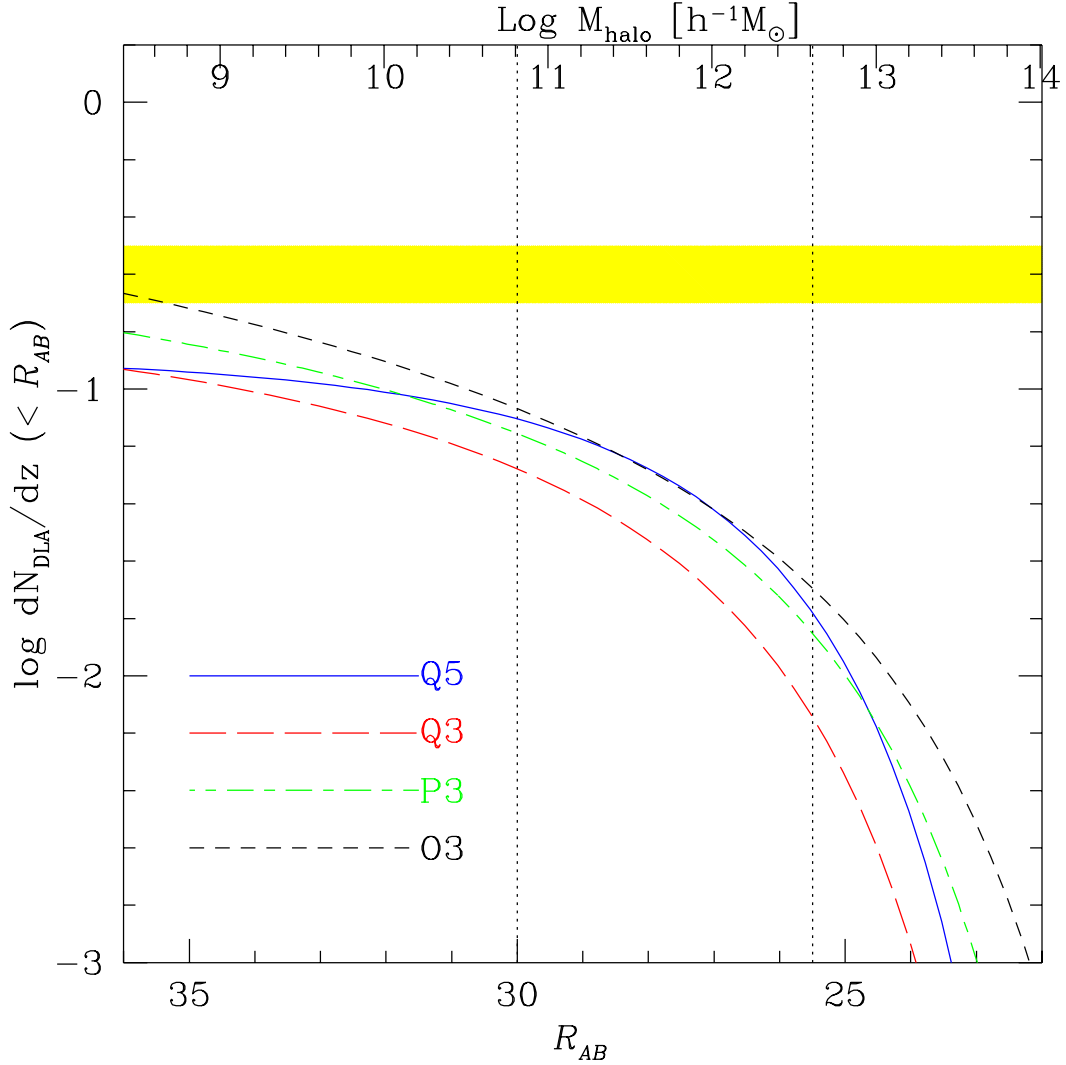}}
\caption{Cumulative distribution function of DLA rate-of-incidence 
as a function of apparent $\Rab$ magnitude. The spectroscopic limit
$\Rab=25.5$ and the limiting magnitude $\Rab=30$ are indicated
by the vertical dotted lines. The top axis gives the corresponding
halo masses using the relationship for the Q3 and Q5 runs; 
$\Rab = -2.5 \log \Mhalo + 57.03 - 5\log h_{70}$ (Eq.~[\ref{eq:Rab}]). 
The yellow band is the observational estimate by 
\citet[][see text for details]{Pro05}. 
 }
\label{fig:cum_abs}
\end{center}
\end{figure}

\begin{figure*}
\begin{center}
\resizebox{9.3cm}{!}{\includegraphics{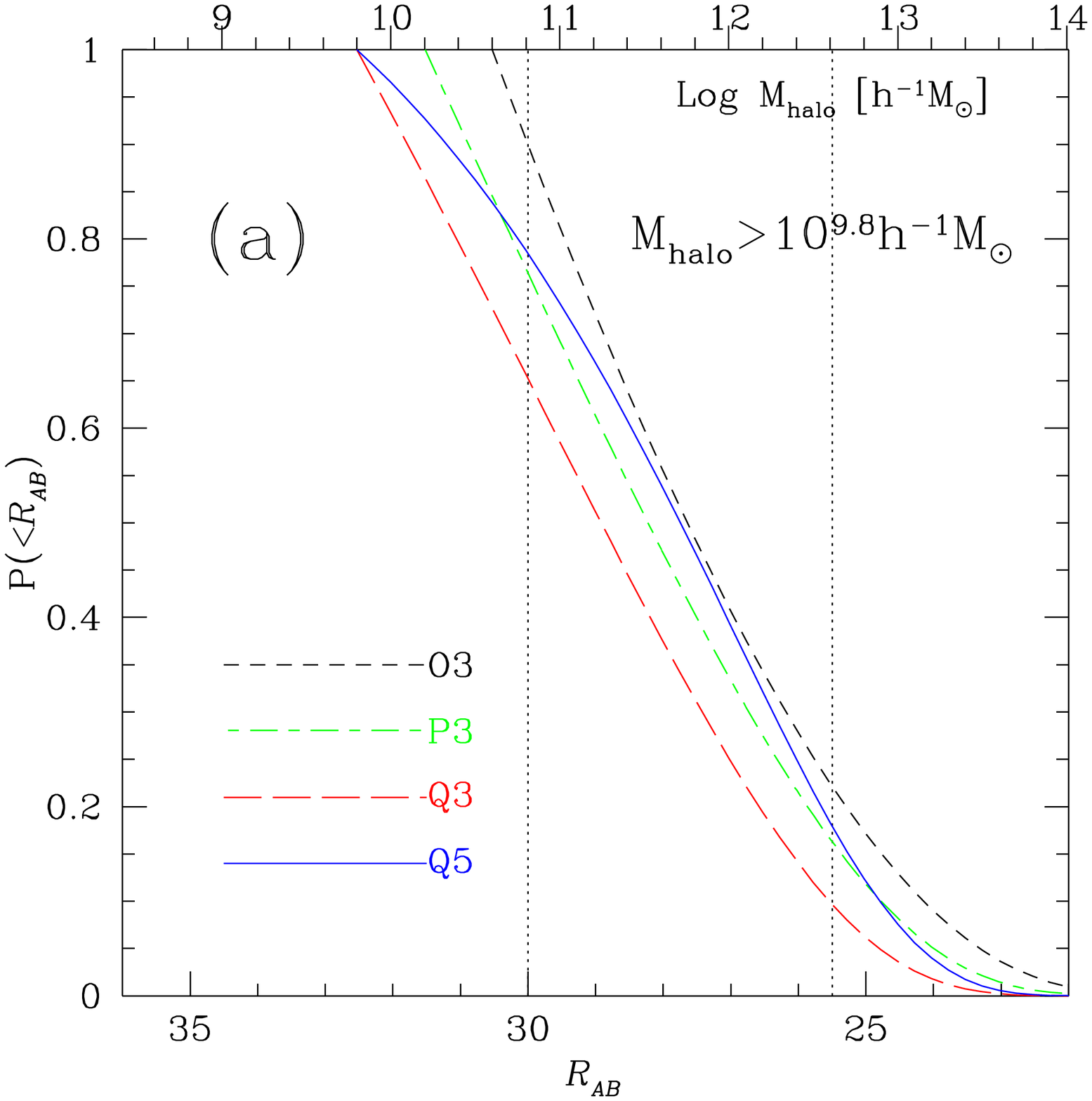}}\\
\vspace{0.3cm}
\resizebox{9.3cm}{!}{\includegraphics{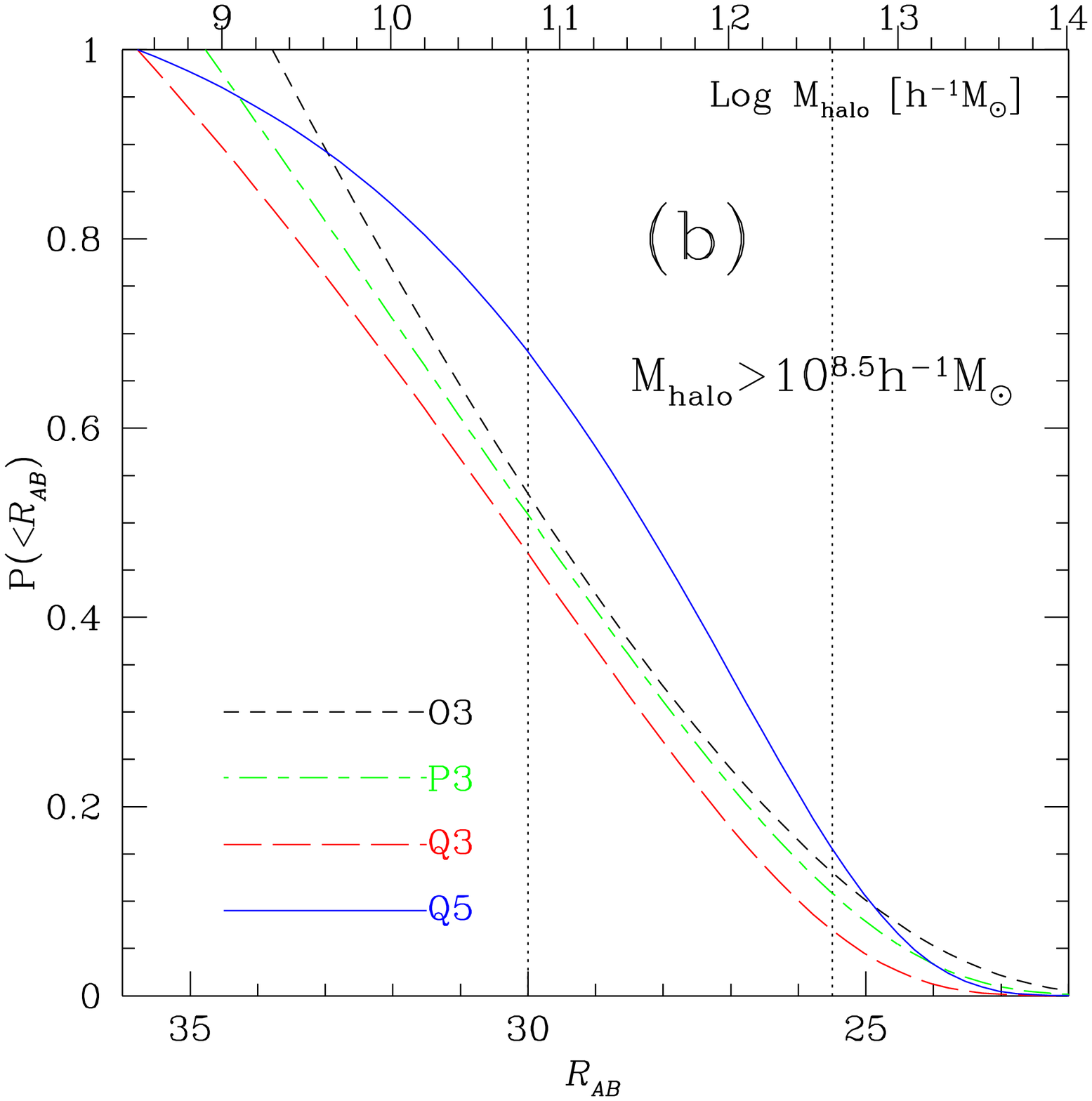}}
\caption{Cumulative probability distribution of DLA rate-of-incidence
as a function of apparent $\Rab$ magnitude at $z=3$. {\it Panel (a)} is when
the distribution is normalized by the value at $\Mhalo = 10^{9.8}\himsun$;
i.e., assuming there would be no DLAs in halos less massive than this value.
This roughly reproduces the result of \citet{Hae00}. 
{\it Panel (b)} is when the distribution is normalized by the value at 
$\Mhalo = 10^{8.5}\himsun$, which is more consistent with the results
of our SPH simulations. The top axis shows the scaling with halo mass 
for the Q3 and Q5 runs, 
$\Rab = -2.5 \log \Mhalo + 57.03 - 5\log h_{70}$ (Eq.~[\ref{eq:Rab}].) 
}
\label{fig:cum_norm}
\end{center}
\end{figure*}

\begin{figure*}
\begin{center}
\resizebox{9.3cm}{!}{\includegraphics{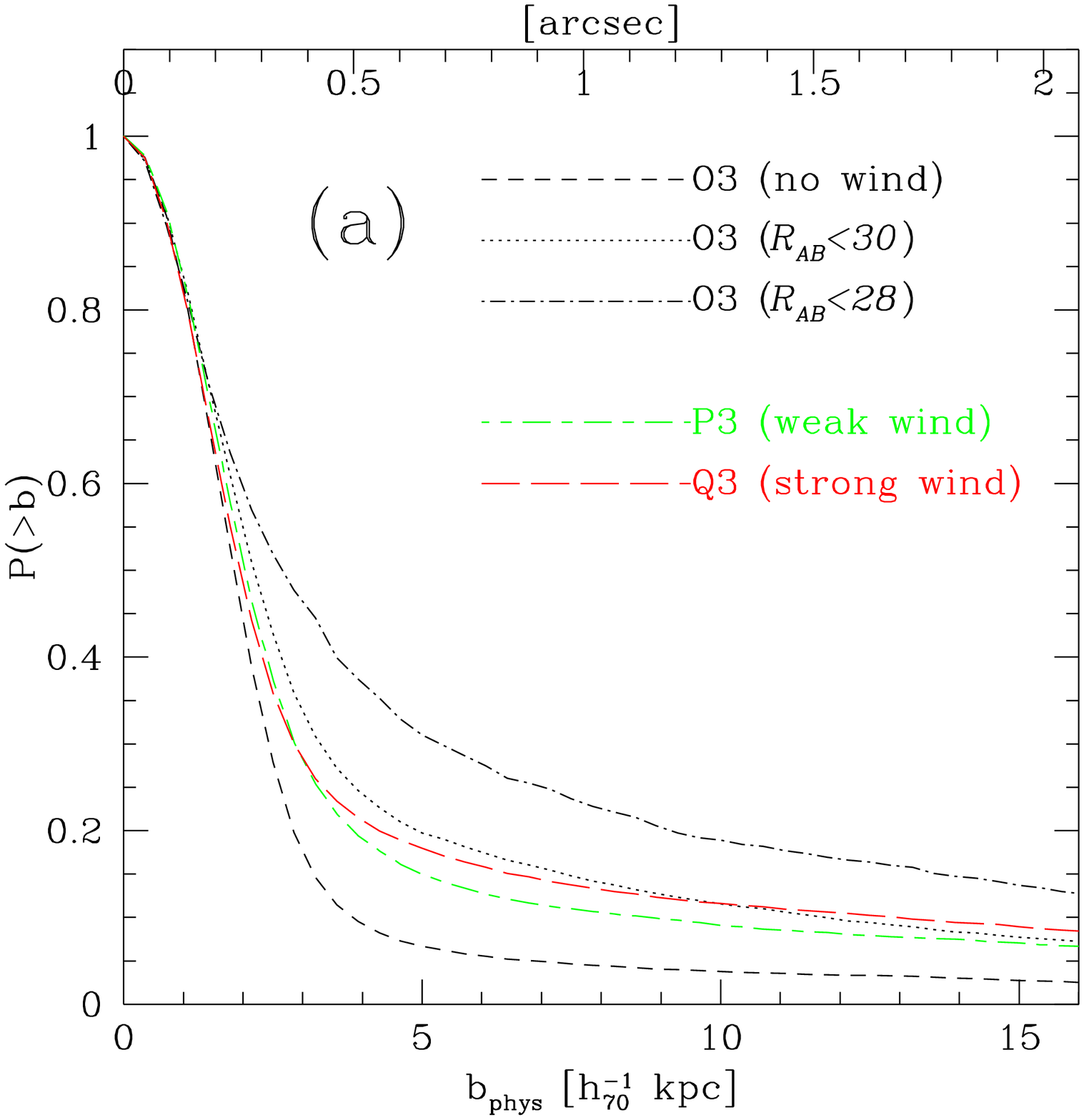}}\\
\vspace{0.3cm}
\resizebox{9.3cm}{!}{\includegraphics{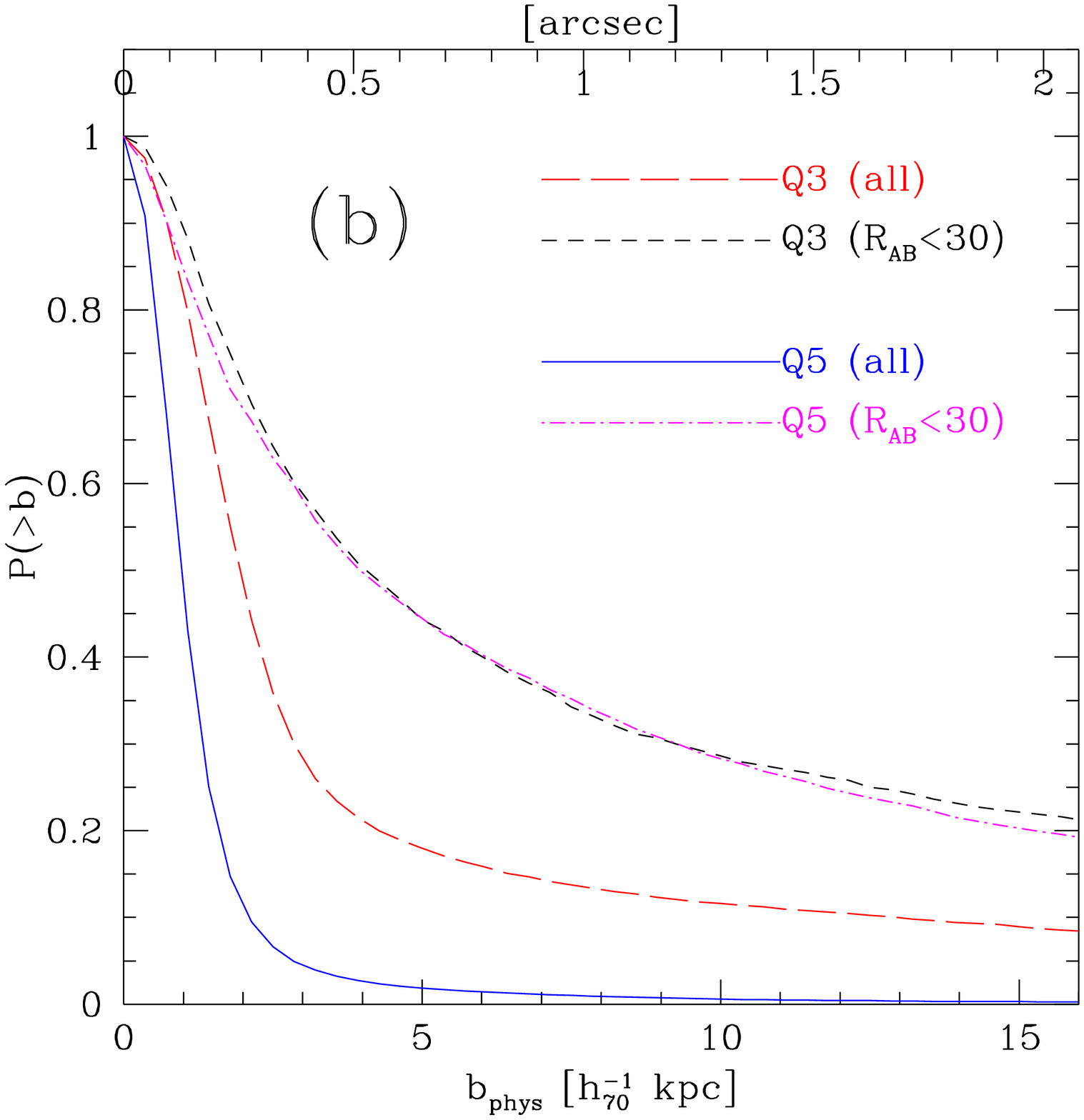}}
\caption{Cumulative probability distribution of DLA
rate-of-incidence as a function of projected impact parameter 
$d_{\rm phys}$ in units of physical $\hsevkpc$ and angular scale 
in units of arc second at $z=3$.
{\it Panel (a)} shows the results of O3 (no wind), P3 (weak wind), 
and Q3 (strong wind) runs. {\it Panel (b)} shows the results of Q3 
and Q5 runs. In both panels, we also show the results when the 
search for the nearest galaxy is limited to those brighter
than a certain $\Rab$ magnitude. }  
\label{fig:impact_norm}
\end{center}
\end{figure*}

\begin{figure}
\begin{center}
\resizebox{15cm}{!}{\includegraphics{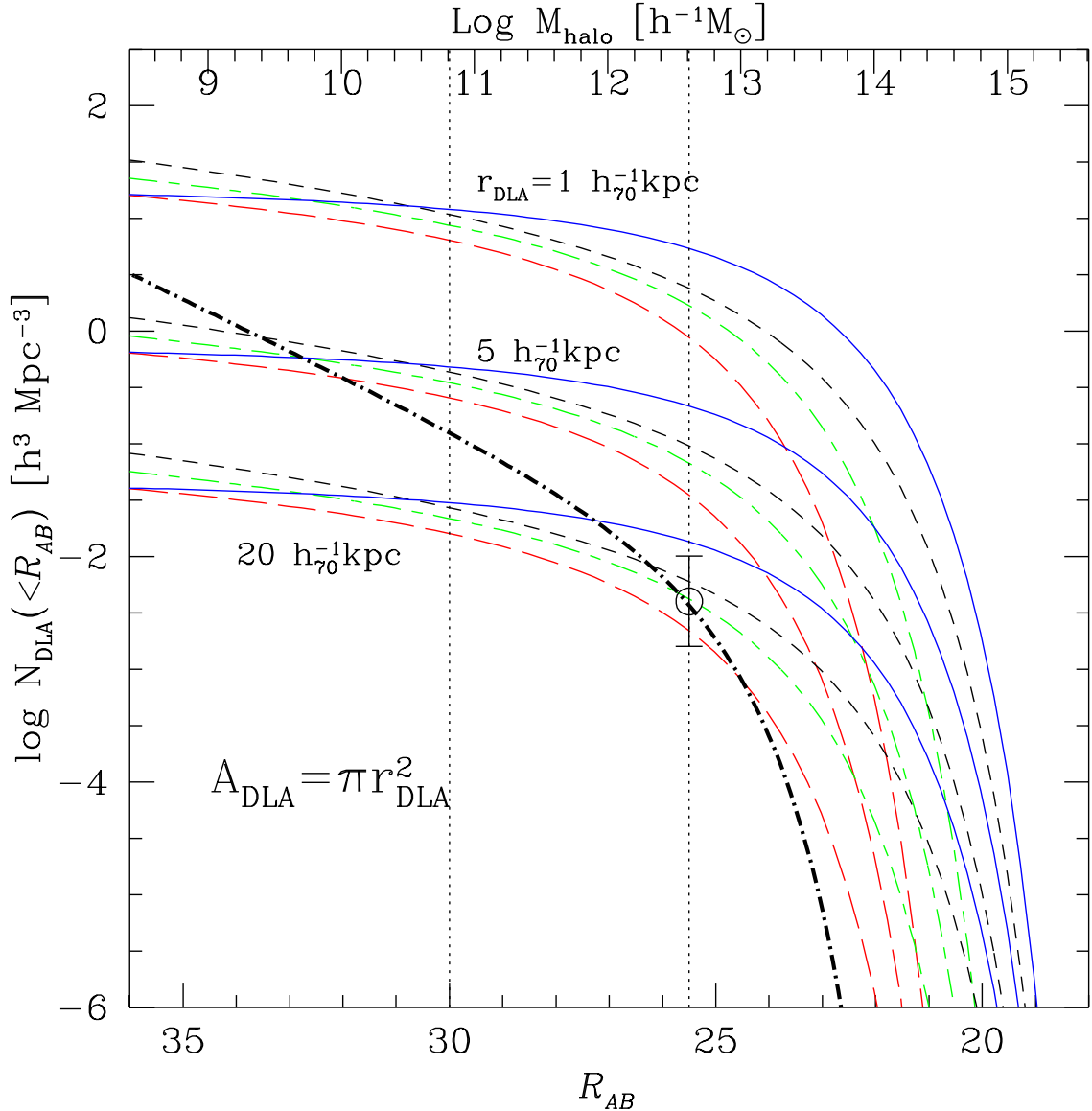}}
\caption{Cumulative, comoving number density of DLAs as a function 
of apparent $\Rab$ magnitude for 3 different values of assumed DLA 
physical radius $\rdla = 1\,\hsevkpc$, 5\,$\hsevkpc$, and 20\,$\hsevkpc$.
The area of each DLA is $A_{\rm DLA} = \pi \rdla^2$. 
For each value of $\rdla$, four different lines are shown: Q5 run
(blue solid line), O3 (black dashed), P3 (green short-dash long-dash), 
and Q3 (red long-dashed). The thick black dot-dashed line is the 
cumulative comoving number density of LBGs obtained by integrating
the observed luminosity function by \citet{Ade00}. The data point
at $\Rab=25.5$ shows the comoving number density of LBGs 
$N_{\rm LBG} =  4\times 10^{-3} h^{-3}\,\mpc^{-3}$. 
}
\label{fig:ndla}
\end{center}
\end{figure}

Finally let us discuss the resolution and convergence issues
of the present work.  
In Figures~\ref{fig:pdf} and \ref{fig:cum_abs} for example, 
the results of the Q3 and Q5 run do not seem to have converged. 
This does not mean that our models for star formation and feedback 
are ill-defined.  
In fact, \citet{SH03a} and \citet{SH03b} showed that 
the code and the simulations converge well in runs with 
star formation and wind feedback. 
The reason for the non-convergence seen in Figures~\ref{fig:pdf} 
and \ref{fig:cum_abs} can be primarily ascribed to the restricted 
dynamic range of current cosmological simulations, 
which is present in almost all modern numerical work. 

In cosmological simulations, the issue of establishing numerical 
convergence is clouded by the fact that an increase of the 
resolution always modifies the problem one solves 
-- one suddenly sees a whole new generation of low-mass galaxies 
that weren't there previously. 
This effect disappears only once the simulations resolve 
all halos capable of cooling, and it was achieved by the `R-series' 
in the earlier work by \citet{NSH04a}, 
in which we deduced our lower limit for the DLA halo mass.  
When lower mass halos are newly resolved in the Q5 run 
which were not present in the Q3 run, 
it steepens the slope of the relation between 
DLA cross-section and halo mass, reducing the relative contribution 
by the low-mass halos to the total DLA incidence.  
This change causes the difference between the results of 
Q3 and Q5 run in Figures~\ref{fig:pdf} and \ref{fig:cum_abs}, 
and it can be regarded as a cautionary sign for the effect 
of resolution on these calculations. 
In the Q5 run, there are 150 dark matter particles for a
halo with $\Mhalo = 10^{8.5}\himsun$, so the results
of the Q5 run should be close to the convergence if not fully converged. 

Note however that we nevertheless find it highly useful to explore 
the consequences and predictions of our particular wind feedback 
model for DLAs, even if the present work eventually can 
only be used to demonstrate the failure of the model.  
The model is numerically well posed, and therefore allows 
various predictions on the properties on DLAs and galaxies 
to be made \citep{NSH04a, NSH04b, NSH05, NSHM, NSHM2, 
Nachos1, Nachos2, Nachos3}.  
We are thus able to determine where our numerical model agrees and 
disagrees with the observational data on DLAs and galaxies.  
It is exactly such a comparison that makes DLA observations 
so useful for theory, because they can teach us in this way about
galaxy formation.  In the future when the 
computers become fast enough to achieve full dynamic range of mass
and spatial scales without employing a series of simulations, 
we will be able to better address the issues of numerical resolution 
and convergence.


\section*{Acknowledgements}

We thank Jeff Cooke and Jason Prochaska for useful discussions. 
This work was supported in part by NSF grants ACI
96-19019, AST 00-71019, AST 02-06299, and AST 03-07690, and NASA ATP
grants NAG5-12140, NAG5-13292, and NAG5-13381.  The simulations were
performed at the Center for Parallel Astrophysical Computing at the
Harvard-Smithsonian Center for Astrophysics.



\begin{deluxetable}{cccccc}
\tablecolumns{6}
\tablewidth{0pc}
\tablecaption{Simulations}
\tablehead{
\colhead{Run} &  \colhead{${N_{\rm p}}$} & \colhead{$m_{\rm DM}$} & \colhead{$m_{\rm gas}$} & \colhead{$\epsilon$} & \colhead{wind}} 
\startdata
O3  & $144^3$ &  $2.42\times 10^7$ & $3.72\times 10^6$ &2.78 & none \cr
P3  & $144^3$ &  $2.42\times 10^7$ & $3.72\times 10^6$ &2.78 & weak \cr
Q3  & $144^3$ &  $2.42\times 10^7$ & $3.72\times 10^6$ &2.78 & strong \cr
Q4  & $216^3$ &  $7.16\times 10^6$ & $1.10\times 10^6$ &1.85 & strong \cr
Q5  & $324^3$ &  $2.12\times 10^6$ & $3.26\times 10^5$ &1.23 & strong \cr
\enddata
\tablecomments{Simulations employed in this study.  All simulations have 
a comoving box size of 10 $\himpc$. The (initial) number of gas particles 
$N_{\rm P}$ is equal to the number of dark matter particles, so the 
total particle count is twice ${N_{\rm p}}$. $m_{\rm DM}$ and 
$m_{\rm gas}$ are the masses of dark matter and gas
particles in units of $\himsun$, respectively, and $\epsilon$ is the
comoving gravitational softening length in units of $\hikpc$,
The value of $\epsilon$ is a measure of spatial resolution.}
\label{table:sim}
\end{deluxetable}

\begin{deluxetable}{cccl}
\tablecolumns{4} \tablewidth{0pc} \tablecaption{Test runs with
different wind parameters} \tablehead{ \colhead{Run} &
\colhead{$\zetadecoup$} & \colhead{$\Lmaxtravel\,[\hikpc]$ }} 
\startdata 
Q3   & 0.1  & 20  & default values \cr 
Q3w0 & 1.0  & 20  & higher $\rhothw$ \cr 
Q3w1 & 0.01 & 20  & lower $\rhothw$ \cr 
Q3w2 & 0.1  & 4   & shorter $\tdecoup$ \cr
Q3w3 & 0.1  & 100 & longer $\tdecoup$
\enddata 
\tablecomments{The density $\rhothw = \zetadecoup\, \rhoth$ denotes 
the threshold density for the decoupling of the hydrodynamic force, 
and $\tdecoup = \Lmaxtravel / v_w$ parameterize 
the maximum allowed time of the decoupling.  }
\label{table:windparams}
\end{deluxetable}

\begin{deluxetable}{ccc}
\tablecolumns{3}
\tablewidth{0pc}
\tablecaption{Fit parameters for test runs}
\tablehead{
\colhead{Run} &  \colhead{slope $\alpha$} & \colhead{normalization $\beta$}} 
\startdata
Q3   & 0.84 & 3.98 \cr
Q3w0 & 0.83 & 3.96 \cr
Q3w1 & 0.80 & 3.99 \cr
Q3w2 & 0.80 & 3.95 \cr
Q3w3 & 0.82 & 3.97 
\enddata
\tablecomments{Fit parameters to the DLA area vs. halo mass relationship. 
}
\label{table:fitparams}
\end{deluxetable}

\begin{deluxetable}{cccccc}
\tablecolumns{6}
\tablewidth{0pc}
\tablecaption{DLA halo masses}
\tablehead{
\colhead{Run} &  \colhead{$\log \Mpeak$} & \colhead{$\log \Mmed$} & \colhead{$\log \Msev$} & \colhead{$\log \Mmean$} & \colhead{$\lgMmean$}
}
\startdata
O3  & (8.5) & 10.1 & 11.1 & 11.8 & 10.4 \cr
P3  &  9.6  & 10.4 & 11.5 & 11.9 & 10.6 \cr
Q3  & 10.8  & 10.7 & 11.7 & 12.0 & 10.8 \cr
Q4  & 11.6  & 11.1 & 12.0 & 12.2 & 11.1 \cr
Q5  & 12.0  & 11.5 & 12.3 & 12.4 & 11.3 \cr
\enddata
\tablecomments{Various DLA halo masses in units of $\himsun$ 
that characterize the 
differential distribution $dN/(dz\,d\log M$) with the distribution 
cutoff at $\Mhalo = 10^{8.5}\himsun$. The following quantities are
listed: peak halo mass $\Mpeak$, median halo mass $\Mmed$ (i.e., 50 
percentile of the distribution), 75 percentile of the distribution 
$\Msev$, mean halo mass $\Mmean$ (Equation~(\ref{eq:mean})), and the 
mean of $\log \Mhalo$ rather than $\Mhalo$ itself for the comparison 
with the result by \citet{Bouche05}. 
The peak halo mass for O3 run is shown in the parenthesis because 
it is the cutoff mass itself of the distribution. }
\label{table:meanmass}
\end{deluxetable}


\end{document}